\documentclass[aps,prd,twocolumn,showpacs,superscriptaddress,groupedaddress]{revtex4-1}  
\usepackage{amssymb}
\usepackage{amsfonts}
\usepackage{amsmath,array}
\usepackage{epsfig}
\usepackage{graphicx}
\usepackage{epstopdf} 
\usepackage{dcolumn}
\usepackage{bm}
\usepackage{lineno}
\usepackage{overpic}
\usepackage{multirow}
\usepackage{color}
\usepackage{bm}
\usepackage[utf8]{inputenc}
\usepackage[perpage,symbol]{footmisc}
\usepackage{subfigure}
\usepackage{enumitem}
\usepackage[colorlinks,
            linkcolor=blue,
            anchorcolor=blue,
            citecolor=blue]{hyperref}
\definecolor{cobalt}{rgb}{0.0, 0.28, 0.67}
\begin{document}

\title{ \Large{New Features in the Electromagnetic Structure of the Neutron}}
\author{
\begin{small}
\begin{center}
M.~Ablikim$^{1}$, M.~N.~Achasov$^{10,c}$, P.~Adlarson$^{67}$, S. ~Ahmed$^{15}$, M.~Albrecht$^{4}$, R.~Aliberti$^{28}$, A.~Amoroso$^{66A,66C}$, Q.~An$^{63,50}$, ~Anita$^{21}$, X.~H.~Bai$^{57}$
% ...
\\
\vspace{0.2cm}
(BESIII Collaboration)\\
\vspace{0.2cm} {\it
$^{1}$ Institute of High Energy Physics, Beijing 100049, People's Republic of China\\
% ...
$^{4}$ Bochum Ruhr-University, D-44780 Bochum, Germany\\
% ...
$^{10}$ G.I. Budker Institute of Nuclear Physics SB RAS (BINP), Novosibirsk 630090, Russia\\
% ...
$^{15}$ Helmholtz Institute Mainz, Johann-Joachim-Becher-Weg 45, D-55099 Mainz, Germany\\
% ...
$^{21}$ Indian Institute of Technology Madras, Chennai 600036, India\\
% ...
$^{28}$ Johannes Gutenberg University of Mainz, Johann-Joachim-Becher-Weg 45, D-55099 Mainz, Germany\\
% ...
$^{50}$ State Key Laboratory of Particle Detection and Electronics, Beijing 100049, Hefei 230026, People's Republic of China\\
% ...
$^{57}$ University of Jinan, Jinan 250022, People's Republic of China\\
% ...
$^{63}$ University of Science and Technology of China, Hefei 230026, People's Republic of China\\
% ...
$^{66}$ (A)University of Turin, I-10125, Turin, Italy; (B)University of Eastern Piedmont, I-15121, Alessandria, Italy; (C)INFN, I-10125, Turin, Italy\\
$^{67}$ Uppsala University, Box 516, SE-75120 Uppsala, Sweden\\
% ...
$^{c}$ Also at the Novosibirsk State University, Novosibirsk, 630090, Russia\\
% ...
}\end{center}
\vspace{0.4cm}
\end{small}
}

\begin{abstract}
Here we present new results for the Born cross section and the effective form factor of the neutron  at the center-of-mass energies ${\bf \sqrt{s}}$ between 2.0 and 3.08 GeV, using 18 data sets corresponding to an integrated luminosity of 647.9 pb${\bf ^{-1}}$ from e${\bf ^+}$e${\bf ^-}$ annihilation reactions collected at the BESIII experiment.
The process $e^{+}e^{-}\to n\bar{n}$ is analyzed with three individual categories to improve the efficiency of $n\bar{n}$ reconstruction.
The cross section of $e^{+}e^{-}\to n\bar{n}$ is measured at 18 c.m. energies where the best precision is 8.1\% at $\sqrt{s}=2.396$ GeV.
The corresponding effective form factors are extracted under the assumption $|G_{E}|=|G_{M}|$.
Our results improve the statistical precision on the neutron form factor by more than a factor of 60 over previous measurements from the FENICE and DM2 experiments and usher in a new era where neutron form factor data from annihilation in the time-like regime is on par with that from electron scattering experiments. In addition, an oscillatory behavior of the effective form factor observed for the proton is discussed for the neutron.
\end{abstract}

\date{\today}


\maketitle

\section{Introduction}
\label{intro}

\begin{figure*}[htb]
	\centering
\hfill\begin{minipage}{0.48\linewidth}
 \hspace{0.5cm}\begin{overpic}[width=0.8\textwidth]{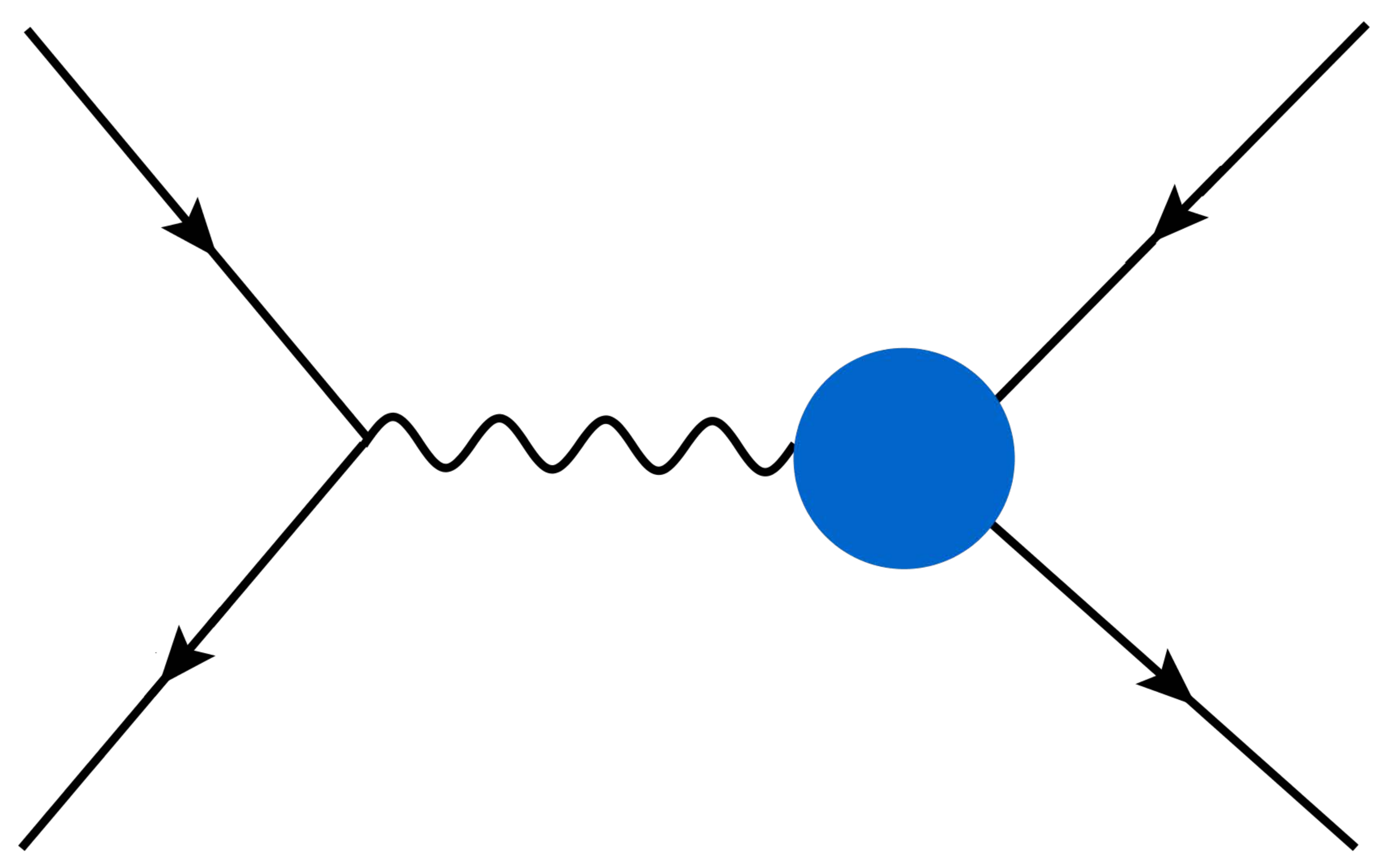}
  \put(16,8){\LARGE{$\rm e^+(k_2)$}}
  \put(16,50){\LARGE{$\rm e^-(k_1)$}}
  \put(77,28){\Large{$\rm J^{\mu}_{had}$}}
  \put(63,8){\LARGE{\color{cobalt}$\rm N(p_2)$}}
  \put(63,50){\LARGE{\color{cobalt}$\rm \bar{N}(p_2)$}}
  \put(8,28){\Large{$\rm j^{\mu}_{lep}$}}
  \put(35,37){\Large{$\rm \gamma^*(q^2)$}}
    \put(-10, 64) {{\bf \Large{(a)}}\color{black}}
  \end{overpic}
\end{minipage}
\begin{minipage}{0.49\linewidth}
	\centering
	  \hspace{0.5cm}\begin{overpic}[width=0.9\textwidth]{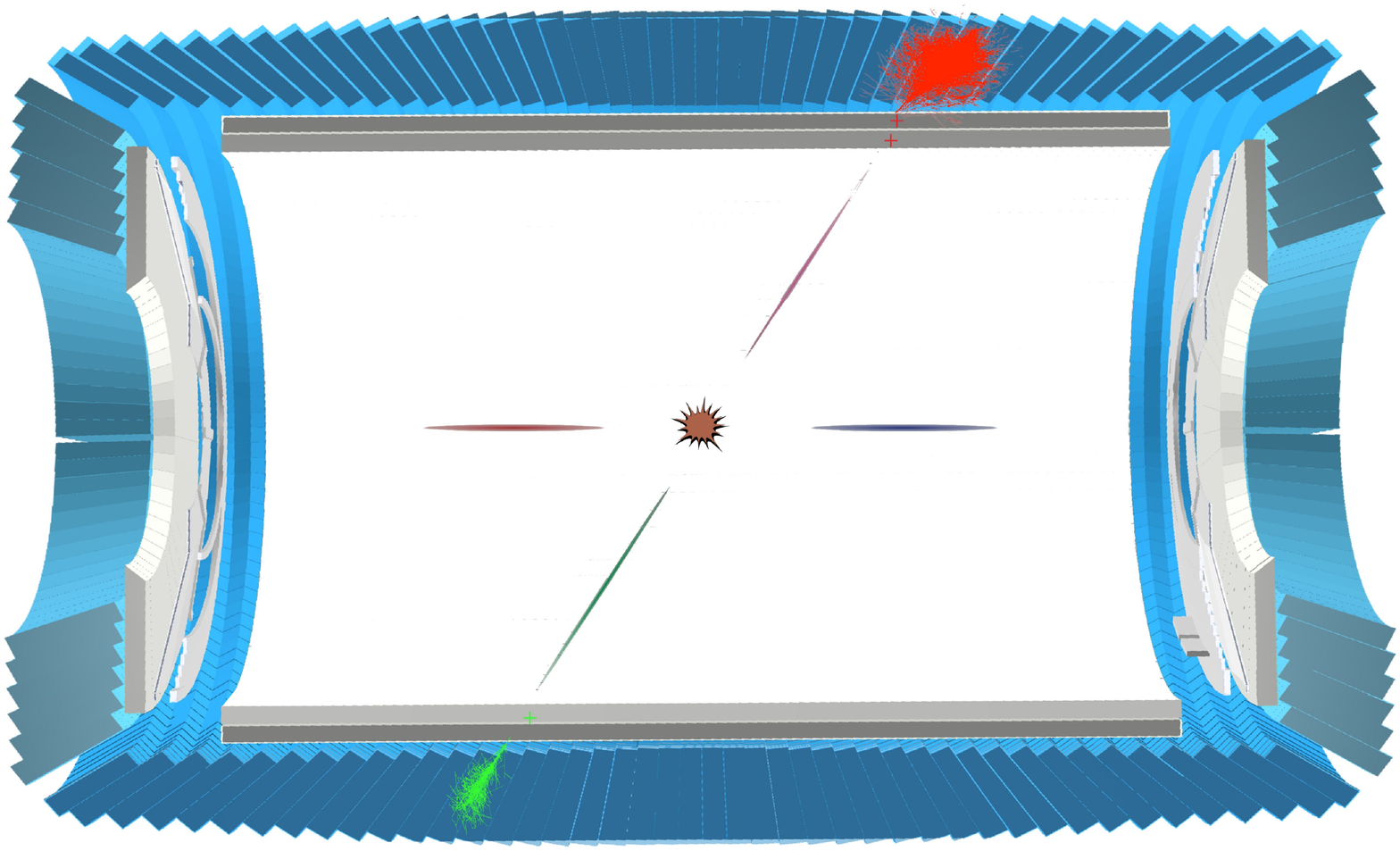}
  \put(-11,28){\Large{$\rm e^+ \color{red}\rightarrow $}}
  \put(97,28){\Large{$\color{blue}\leftarrow \color{black}\rm e^-$}}
  \put(58,55){\small{ \color{red} $\bigstar$ }}
  \put(34,17){\small{ \color{green} $\bigstar$ }}
  \put(32,20){\LARGE{\color{cobalt}$\rm n$}}
  \put(65,50){\LARGE{\color{cobalt}$\rm \bar{n}$}}
  \put(-10, 60) {{\bf \Large{(b)}}\color{black}}
     \end{overpic}
\end{minipage}
\hfill
	\caption{(a) The lowest order Feynman diagram for the process $e^+e^-\to N \bar{N}$. (b) Typical response in the BESIII detector for the signal process $e^+e^-\to n\bar{n}$ shown in the parallel plane to the electron-positron beam direction. The red (green) stars and showers represent the detector response of the signal process in the Time-of-Flight system and the Electromagnetic Calorimeter for the anti-neutron (neutron).}
	\label{fig:signal}
\end{figure*}

The neutron is a bound system of three valence quarks and a neutral sea consisting of gluons and quark-antiquark pairs. Although the proton was discovered in 1919 and the neutron in 1932, the structure of the nucleon is still not fully understood. Over the years, investigations of the nucleon raised new questions in experiments and theory, such as the spin crisis \cite{spincrisis} and the mass decomposition \cite{masscrisis}. One famous example in scattering experiments is the proton radius puzzle, showing a discrepancy of $2.7\sigma$ between measurements with muonic and electronic hydrogen \cite{protonradius,protonradius2,protonradius3} with some remaining mysteries still to be clarified by future experiments. Another controversy appeared around the remarkably different charge density of the neutron among various models, which reveal an opposite sign of the mean-square of charge radius \cite{neutronradius1,neutronradius2}. In annihilation experiments, a long-standing puzzle arose with the results from electron-positron annihilation reported by the FENICE \cite{FENICE} and DM2 \cite{DM2} experiments, indicating a stronger photon-neutron interaction than the corresponding interaction with a proton. This observation is difficult to reconcile with theoretical expectations \cite{Chernyak,rnp1}. A recent example is an interesting oscillating behavior observed in the proton form factor in a measurement by the BaBar experiment \cite{osz1}, indicating a complex structure in the effective form factor data. These open questions might be answered through the measurement of observables like the Born cross section $\sigma_{B}(q^2) \equiv \sigma_{B}$, the corresponding effective form factor $|G(q^2)| \equiv |G|$, the electric $G_E(q^2) \equiv G_E$ and magnetic $G_M(q^2) \equiv G_M$ form factors of the nucleon.  These form factors are all functions of the squared momentum transfer, $q^2 = (k_1 + k_2)^2  = (p_1 + p_2)^2= s$, where $k_1,\ k_2,\ p_1,\ p_2$ are the incoming and outgoing four-momenta of the (anti-)~lepton and (anti-)~nucleon, respectively. The form factors parameterize the coupling of a virtual photon $\gamma^*(q^2)$ with the hadronic current $J^{\mu}_{had}$. For the electron-positron annihilation process into a nucleon-anti-nucleon pair, as described by the leading order Feynman diagram in Fig.~\ref{fig:signal}(a), $\sigma_{B}$ and $|G|$ are defined as:

\begin{equation}\footnotesize
\begin{split}
\sigma_{B} = \frac{4\pi\alpha_{em}^2\beta C(q^2)}{3q^2}\left[|G_M(q^2)|^2 + |G_E (q^2)|^2\frac{1}{2\tau} \right], \\
|G| = \sqrt{\frac{2\tau |G_M(q^2)|^2 + |G_E(q^2)|^2}{2\tau +1}}
\end{split}
\end{equation}
Here, $\alpha_{em}$ is the electromagnetic fine structure constant, $\beta$ is the velocity of the final state nucleon or anti-nucleon, $\tau = q^2/4m_N^2$ with $m_N$ the nucleon mass, and $C(q^2)$ is the S-wave Sommerfeld-Gamow factor for the Coulomb correction \cite{coulomb2}, which is equal to 1 for neutral baryons.

The Beijing Electron-Positron Collider II is a symmetric electron-positron collider, operating in the center-of-mass~(c.m.) energy region between 2.0 and 4.6 GeV. We study neutron and anti-neutron pairs produced in $ e^+e^-$ annihilations for $\sqrt{s}$ between 2.0 and 3.08~GeV. This data set represents the first high luminosity off-resonance energy scan, which enables us to perform a precise measurement of $\sigma_{B}$ and $|G|$ for the process $e^+e^-\to n \bar{n}$ at 18 c.m.~energies. The Beijing Spectrometer III (BESIII) experiment \cite{besiii} at the collider has been optimised for the reconstruction of charged particles and photons using the Main Drift Chamber (MDC) inner tracker to measure momenta and an Electromagnetic Calorimeter (EMC) constructed with CsI(Tl) crystals to measure energy deposition. A Time-of-Flight (TOF) system consisting of plastic scintillator bars outside of the drift chamber measures the flight time of charged particles. In addition, a method for the flight time measurement for neutral particles has been developed for this analysis, as introduced in the section \textbf{Appendix}~\ref{app:rec}. A Muon Counter system (MUC) is used to identify muons and to reject cosmic ray background. The analysis of $e^+e^-\to n \bar{n}$ with the BESIII detector is very challenging due to the reconstruction of the two neutral hadrons in the final state in the absence of a hadronic calorimeter and the need for a corresponding efficiency calibration. A schematic representation of the BESIII detector with a typical response from the signal process $e^+e^-\to n\bar{n}$ is shown in Fig. \ref{fig:signal}(b).

\section{Analysis}

To maximize the reconstruction efficiency, the data are classified into three sub-sets (i = A,B,C) depending on the interaction of the signal particles with the detector. Events with signals from a knockoff proton interaction in the TOF plastic scintillators and associated corresponding hadronic showers measured with the EMC from both the neutron and anti-neutron are classified as category A. Events with showers in the EMC from both particles, but only a measured knockoff proton interaction from the anti-neutron are assigned to category B. Events lacking any TOF interaction but with reconstructed hadronic showers measured in the EMC from both signal particles are classified as category C. Every signal event belongs to only one category, and use of all three categories guarantees a high efficiency of the signal reconstruction. We combine the statistically independent results from the three categories using inverse-variance weighting. More details on the signal reconstruction procedures are given in the section \textbf{Appendix}~\ref{app:sel}.

\begin{figure}[htb]
	\centering
  \begin{overpic}[width=0.48\textwidth]{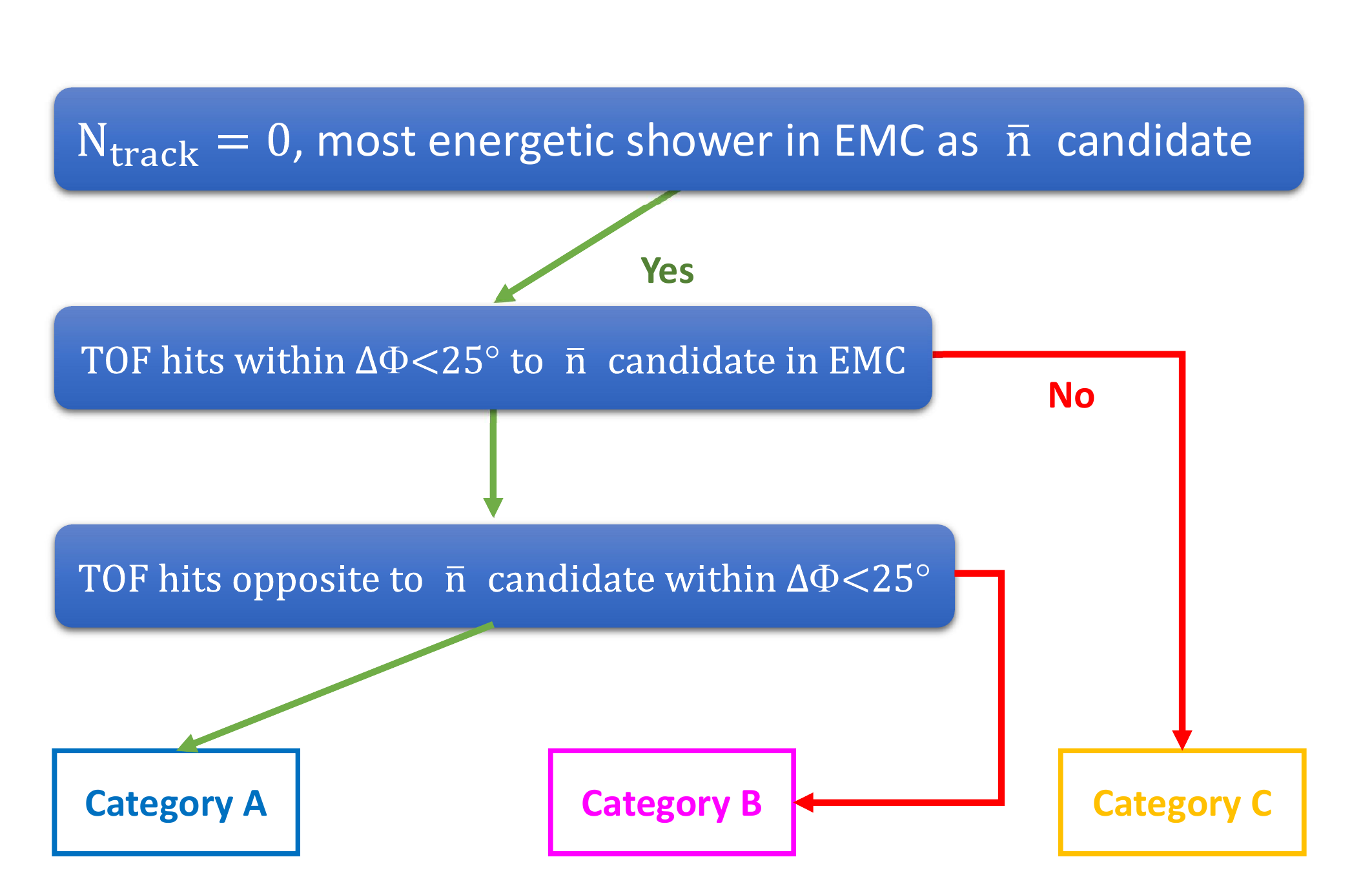}
     \end{overpic}
	\caption{Classification logic for the signal selection. The signal process is reconstructed from events without charged tracks. The most energetic EMC response in the event is identified as from the antineutron. If a TOF response can be matched to the antineutron response in the EMC, the event is classified as category A or B, else as C. If a TOF response is reconstructed at the opposite side of the detector with respect to the antineutron in the EMC, the event is categorized as A, else as B. }
	\label{fig:classification}
\end{figure}

\begin{figure*}[htb]
	\centering
  \begin{overpic}[width=5.5cm,height=4.2cm]{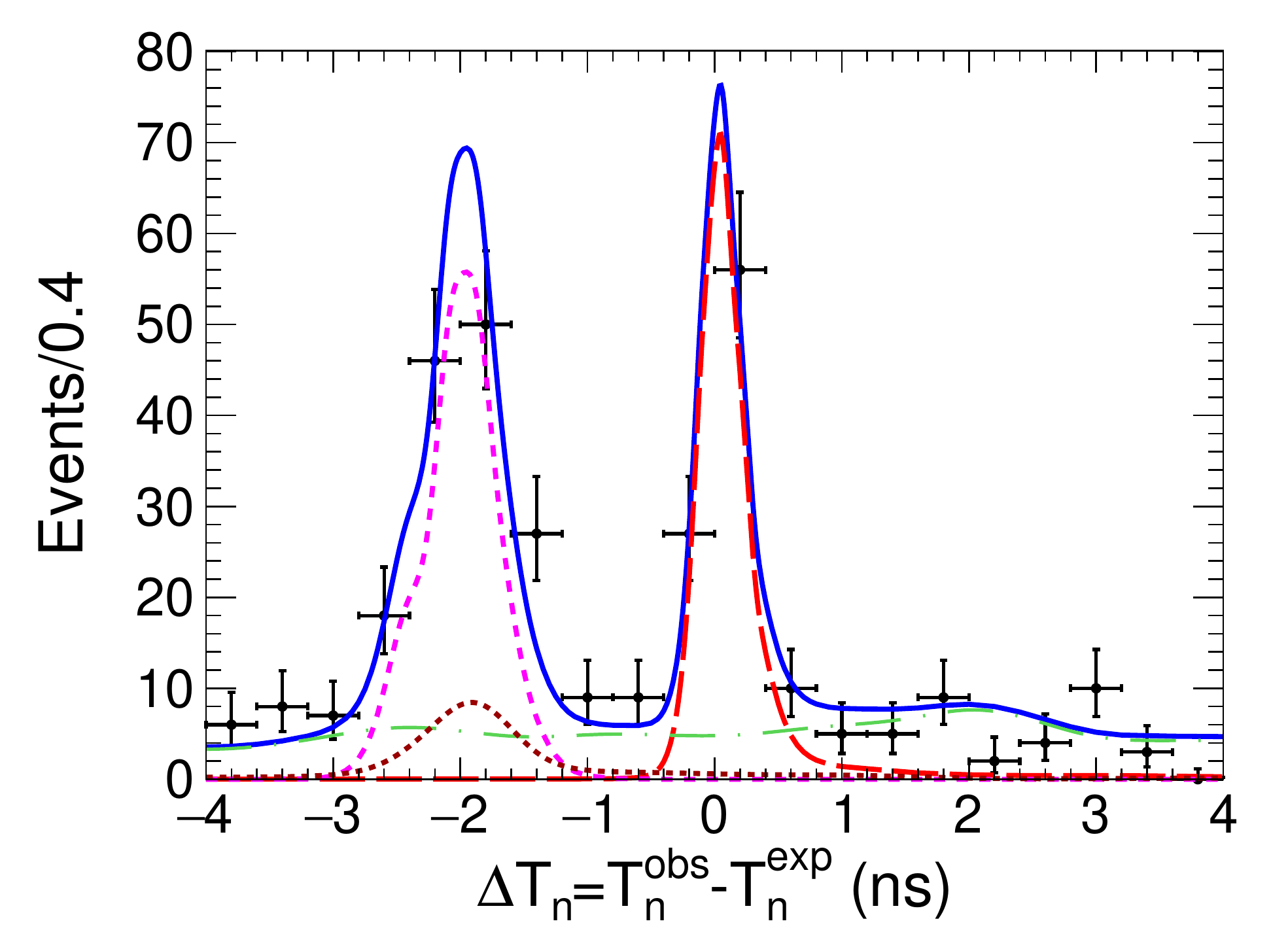}
     \put(21, 60) {\color{red}\Large{A}\color{black}}
     \end{overpic}
  \begin{overpic}[width=5.5cm,height=4.2cm]{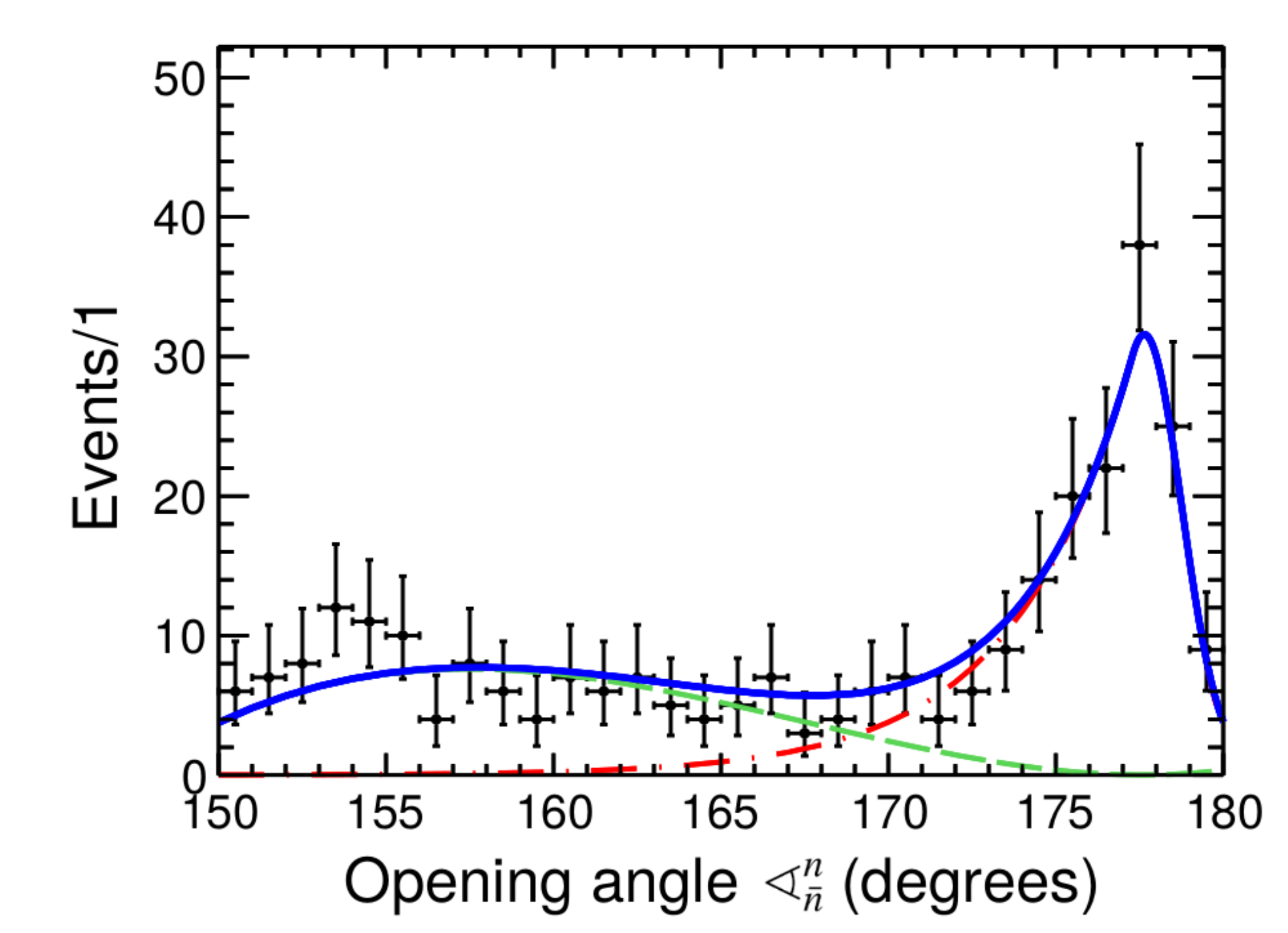}
     \put(21, 60) {\color{red}\Large{B}\color{black}}
     \end{overpic}
  \begin{overpic}[width=5.5cm,height=4.2cm]{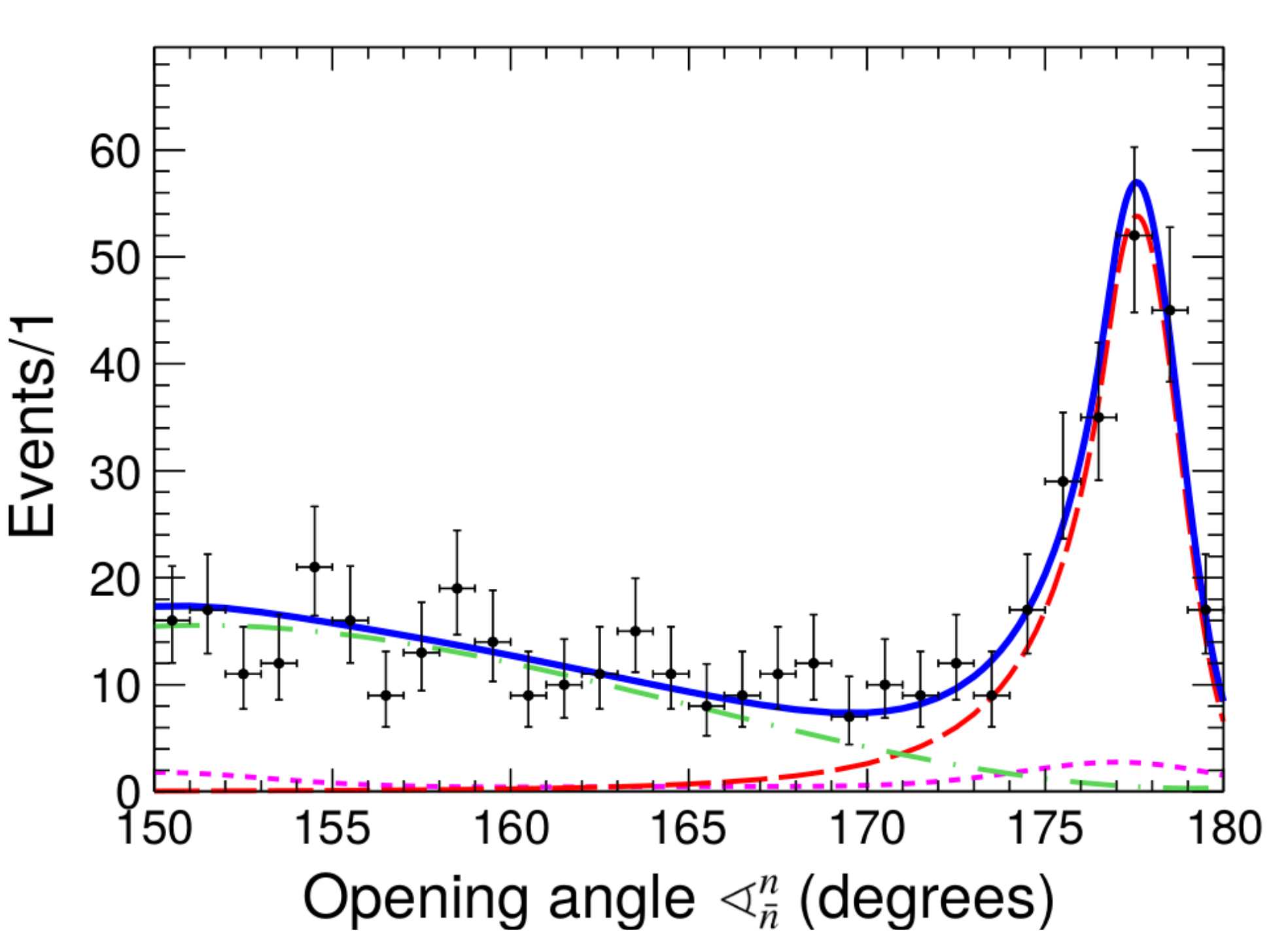}
     \put(16, 60) {\color{red}\Large{C}\color{black}}
     \end{overpic}
	\caption{Examples of the extraction of reconstructed signal events $N_s$ from the pre-selected data sample at the c.m. energy $\sqrt{s}=2.396$~GeV. The reconstructed events of three sub-sets of the data are presented
depending on the corresponding detector impact as a function of (A) $\Delta T_n$
and (B, C) $\sphericalangle^{n}_{\bar{n}}$, where $\Delta T_n$ is the difference between the measured and expected time-of-flight of the neutron in the TOF, $\sphericalangle^{n}_{\bar{n}}$ is opening angle between the anti-neutron and neutron showers in the EMC. The logic for the classification into the three sub-sets of data is shown in Fig.~\ref{fig:classification}.
Black dots with errors indicate BESIII data. The red, green, pink, brown, and blue lines show the signal-, beam-related and cosmic ray-, multi-hadronic-, and di-gamma background components, and the total fit based on Eq.~(\ref{eq:fit}), respectively. The uncertainties shown are statistical only.}
	\label{fig:extraction}
\end{figure*}

The data that pass the signal selection still contain non-negligible background contributions coming mostly from multi-hadronic processes, beam-associated processes and cosmic rays, as shown in the Fig. \ref{fig:extraction}. We investigate the amount and distributions of the remaining background events with dedicated Monte Carlo (MC) simulated events that mimic the detector response for various electrodynamic ($i.e.$ di-gamma) and multi-hadronic processes. Background from cosmic rays and the beam-associated backgrounds including interactions between the beam and the beam pipe, beam and residual gas, and the Touschek effect \cite{besiiibg}, are studied with two data samples collected when the electron and positron beams were not in the collision mode. The number of genuine signal events $\mathcal{N}^s_i$ is extracted from the data samples by fitting to the following distributions: for category A signal events we use the difference $\Delta T_n $ between the time-of-flight of the neutron measured with TOF and the calculated expected flight time; for signal events in categories B and C, the fit is applied to the opening angle $\sphericalangle_{\bar{n}}^{n}$ between the positions of neutron and anti-neutron measured in the Electromagnetic Calorimeter assuming that they originated from the $e^+e^-$ collision point.
An unbinned maximum likelihood fit is performed to determine the $\mathcal{N}^s_i$.
The likelihood function $\mathcal{F}_i$ is constructed by a set of probability density functions~($PDF$) for the signal ($PDF^s_{i}$) and background ($PDF^b_{i}$) contributions and characterized by either $\Delta T_n $ or $\sphericalangle_{\bar{n}}^{n}$. To model the signal event distribution $PDF^s_{i}$, we use MC simulated samples of signal process events generated with {\sc conexc}~\cite{KKMC}.
Specifically:

\begin{equation}
\label{eq:fit}
\centering
\begin{split}
\mathcal{F}_i\left[ obs \right] = \mathcal{N}^s_{i} \cdot PDF^s_{i} + \sum_b \mathcal{N}^b_{i} \cdot PDF^b_{i} \hfill \\
obs=\Delta T_n(i=A)\ or\  \sphericalangle_{\bar{n}}^{n}(i=B,C)
\end{split}
\end{equation}
where {\it{b}} indicates the \textrm{beam-associated,\ multi-hadronic} $e^+e^-\to\gamma\gamma$ backgrounds.
Details for the MC simulation could be found in {\bf Appendix}~\ref{app:mc}.

The reconstruction efficiency, $\varepsilon_{i}$, for the signal process is determined from the exclusive signal MC simulation as well as from data and the additional MC simulation for the physics processes: $e^+e^-  \to J/\psi \to p\bar{n}\pi^-,\ e^+e^-  \to J/\psi \to \bar{p}n\pi^+,\ e^+e^-\to p\bar{p}$, and $e^+e^-\to \gamma\gamma$. Using these samples we correct the differences of the detector response between the data and signal MC simulation. Details for the correction of the signal reconstruction efficiency are provided in the section  {\bf Appendix}~\ref{app:results}.
Using a precisely measured integrated luminosity $\mathcal{L}_{\rm int}$ \cite{luminosity},
the cross section $\sigma_{B}^i$ (approximated as Born cross section) corrected for initial state radiation
and vacuum polarization $(1+\delta )_i$, and the corresponding form factor $|G^i|$ are determined for each classification category as:

\begin{equation}\label{formulas}
\centering
\begin{split}
\sigma^i_{B} = \frac{\mathcal{N}^s_{i}}{\mathcal{L}_{\rm int}\varepsilon_{i}(1+\delta)_i}, \\
|G^i| = \sqrt{\frac{\sigma^i_B}{\frac{4\pi\alpha_{em}^2\beta}{3q^2}\left(1+\frac{1}{2\tau} \right)}}
\end{split}
\end{equation}

\noindent Here $\mathcal{L}_{\rm int}$ is the measured integrated luminosity \cite{luminosity} and  $(1+\delta )_i$ is the initial state radiation and vacuum polarization correction.

\section{Results}
The results from category A, B, and C are consistent with each other within one standard deviation at all c.m.~energies, as shown in the Appendix Fig. \ref{c9:fig:bornsum}. We use inverse-variance weighting to combine these individual results to reduce the statistical uncertainty using the following expressions:

\vspace{-0.3cm}
\begin{equation}
\label{eqn:weightedXS1}
\begin{split}
\sigma_{B} = \sum_i w_{i}\sigma_B^{i} , (i=\textrm{A,\ B,\ C}) \\
 \Delta\sigma_{B} = \sqrt{\frac{1}{\sum_i \sum_j W_{i,j}}} , \\
w_i = \frac{\sum_j W_{i,j}}{\sum_i \sum_j W_{i,j}} , \\
W = [\boldsymbol{\Delta\sigma^T  \rho \Delta\sigma} ]^{-1}
\end{split}
\end{equation}

\begin{figure*}[htb]\label{fig:results}
	\centering
  \begin{overpic}[width=0.49\textwidth]{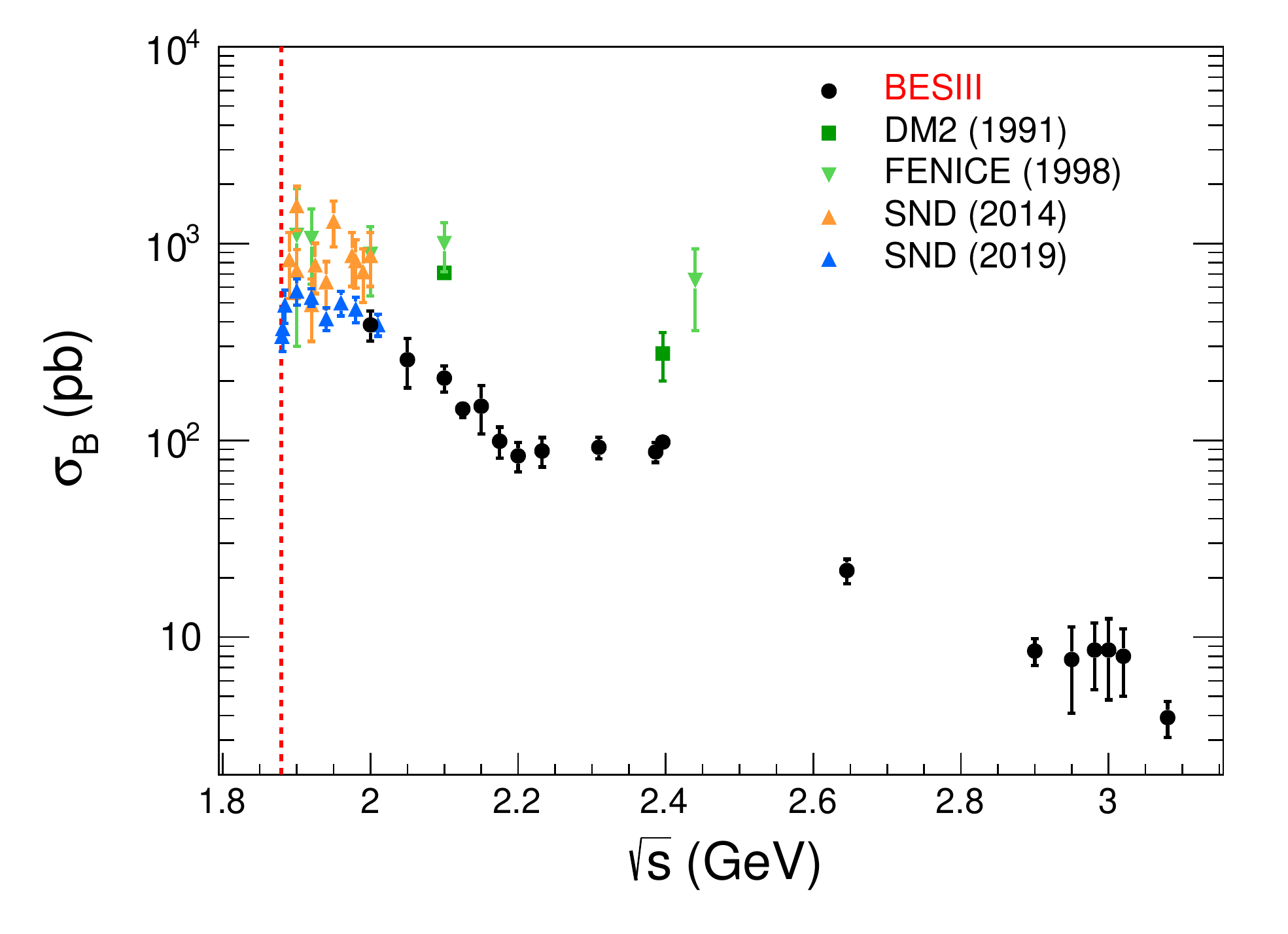}
     \put(-2, 66) {{\bf \Large{(a)}}\color{black}}
     \end{overpic}
     \begin{overpic}[width=0.49\textwidth]{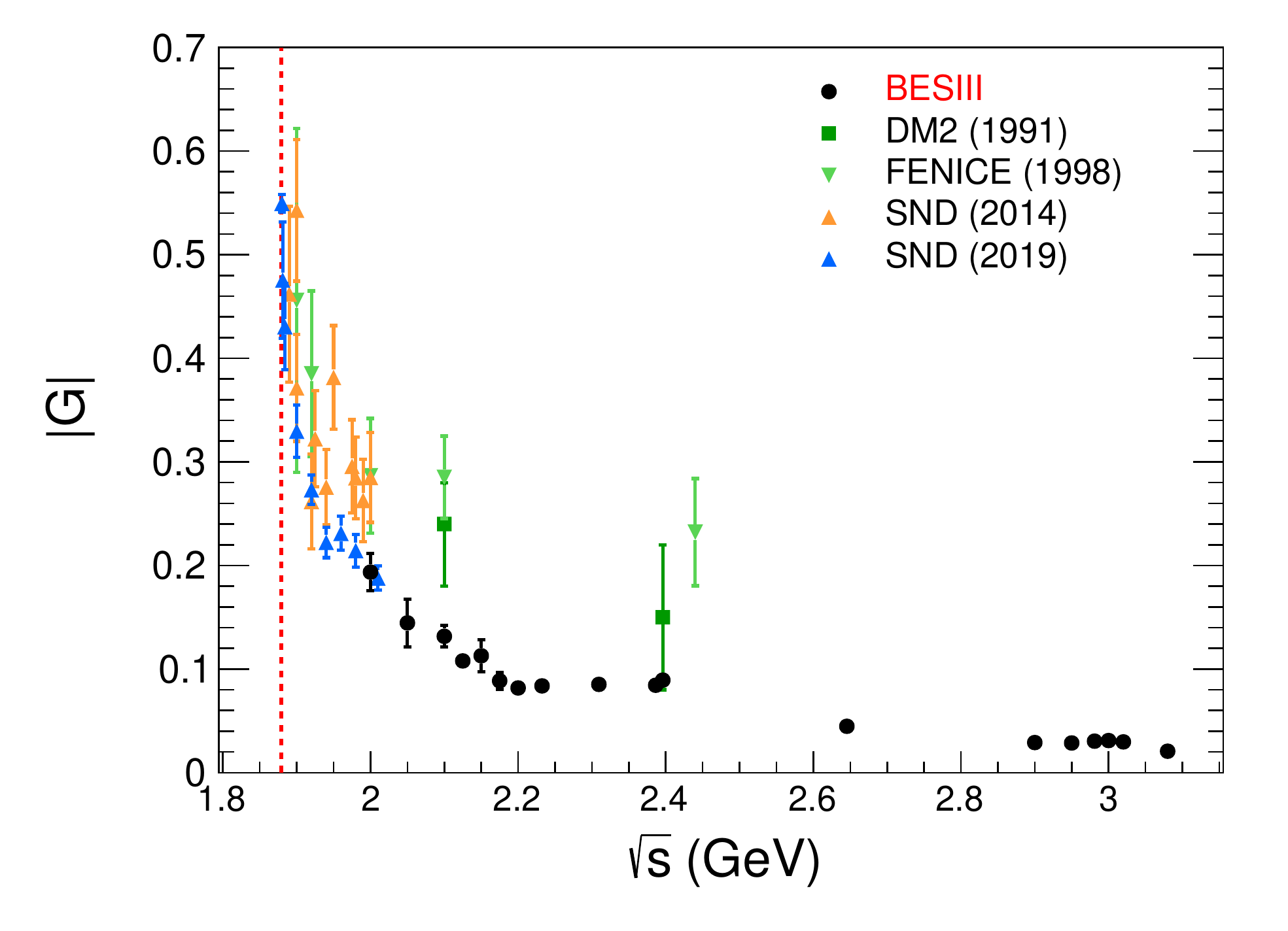}
     \put(-2, 66) {{\bf \Large{(b)}}\color{black}}
     \end{overpic}
	\caption{Results for (a) the Born cross section $\sigma_B$ and (b) the corresponding form factor $|G|$ for the process $e^+e^-\to n\bar{n}$ . The black solid circles are the BESIII results from this analysis. Existing results from the FENICE \cite{FENICE}, DM2\cite{DM2}, and SND \ \cite{snd2014,snd2017} experiments  are shown as green triangles, green squares, light and dark blue triangles, respectively. The red dashed line indicates the production threshold for the signal process. The total uncertainties shown are a sum in quadrature of the statistical and systematic uncertainties.  }
\end{figure*}

The extracted results at 18 c.m.~energies are listed in Table \ref{tab:results} and shown in Fig.~\ref{fig:results}. As sources of systematic uncertainty, we consider category-specific sources, as well as those that are common to two or more categories which introduce correlations. More details on the signal extraction and the evaluation of the systematic uncertainty sources are provided in section \textbf{Appendix}~\ref{app:results}.

\begin{table}[!h]
\footnotesize
\tabcolsep=0.12cm
\begin{center}
\caption{The Born cross section $\sigma_{B}$ for the process $e^+e^-\to n\bar{n}$ and the corresponding $|G|$ of the neutron. The first uncertainty is statistical and the second one systematic. }
\begin{tabular}{c|c|c|c}
\hline
\hline
$\sqrt{s}$ (GeV) & $\mathcal{L}_{\rm int}$ (pb$^{-1}$) & $\sigma_{B}$ (pb) & $|G|$ ($\times10^{-2}$) \\
\hline
2.0000 &  10.1  & $386\pm55\pm37$      &$19.0\pm1.3\pm0.9$  \\
2.0500 &  3.34  & $256\pm67\pm16$      &$14.8\pm1.9\pm0.5$  \\
2.1000 &  12.2  & $207\pm24\pm19$      &$13.0\pm0.8\pm0.6$  \\
2.1250 &   108  & $145\pm6\pm12$       &$10.8\pm0.2\pm0.4$  \\
2.1500 &  2.84  & $149\pm38\pm12$      &$10.9\pm1.4\pm0.4$  \\
2.1750 &  10.6  & $99\pm16\pm8$        &$ 8.8\pm0.7\pm0.4$  \\
2.2000 &  13.7  & $83\pm12\pm6$        &$ 8.1\pm0.6\pm0.3$  \\
2.2324 &  11.9  & $88\pm13\pm7$        &$ 8.3\pm0.6\pm0.3$  \\
2.3094 &  21.1  & $93\pm 9\pm7$        &$ 8.6\pm0.4\pm0.3$  \\
2.3864  &   22.5 & $87\pm8\pm6$&  $8.4\pm0.4\pm0.3$ \\
2.3960  &   66.9 & $98\pm5\pm6$&  $8.9\pm0.2\pm0.3$ \\
2.6454  &   67.7 & $22\pm2\pm2$&  $4.5\pm0.2\pm0.2$ \\
2.9000  &    105 & $ 8.5\pm1.1\pm0.7$&  $3.0\pm0.2\pm0.1$ \\
2.9500  &   15.9 & $ 7.7\pm2.9\pm1.0$&  $2.9\pm0.5\pm0.2$ \\
2.9810  &   16.1 & $ 8.6\pm2.9\pm1.0$&  $3.1\pm0.5\pm0.2$ \\
3.0000  &   15.9 & $ 8.6\pm3.4\pm1.4$&  $3.1\pm0.6\pm0.2$ \\
3.0200  &   17.3 & $ 8.0\pm2.8\pm1.0$&  $3.0\pm0.5\pm0.2$ \\
3.0800  &    126 & $ 3.9\pm0.7\pm0.5$&  $2.1\pm0.2\pm0.1$ \\
\hline
\hline
\end{tabular}
\label{tab:results}
\end{center}
\end{table}

\begin{figure*}[htb]
	\centering
     \begin{overpic}[width=0.49\textwidth]{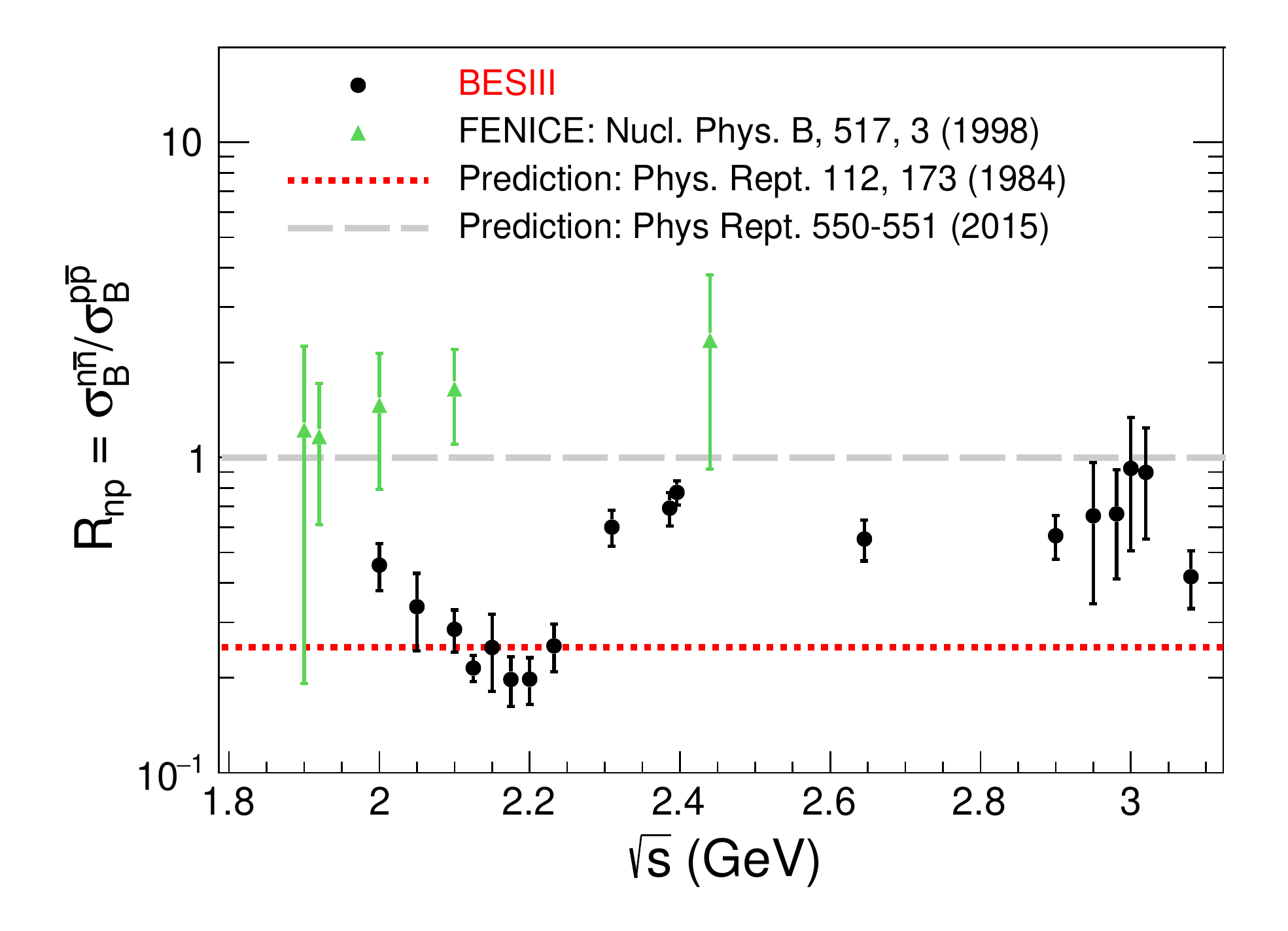}
     \put(-2, 66) {{\bf \Large{(a)}}\color{black}}
     \put(81.5, 61.6) {\tiny {\cite{FENICE}}}
     \put(84, 57.8) {\tiny {\cite{Chernyak}}}
     \put(81.9, 54.) {\tiny {\cite{rnp1}}}
     \end{overpic}
     \begin{overpic}[width=0.49\textwidth]{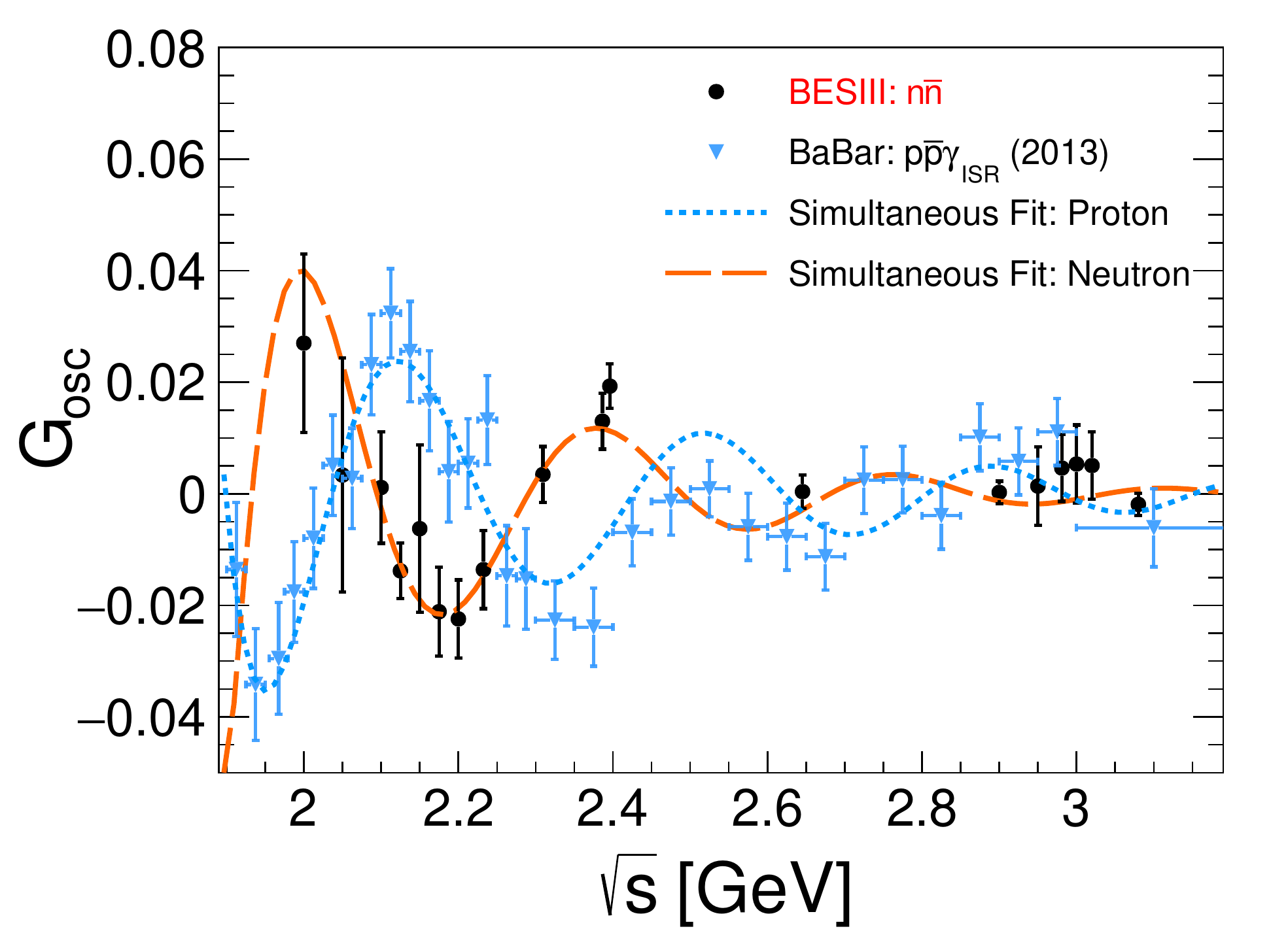}
      \put(-2, 66) {{\bf \Large{(b)}}\color{black}}
       \put(87., 59.6) {\tiny {\cite{osz1}}}
     \end{overpic}
	\caption{(a) The ratio $R_{np}$ is shown using the results from this analysis and the existing data on the proton from~\cite{protonscan} together with the results from the FENICE experiment~\cite{FENICE}. The red fine dashed and grey wide dashed lines are two predictions from~ \cite{Chernyak} and~\cite{rnp1}. (b) Fit to the deviation of the effective form factor $|G|$ of the nucleon from the dipole law. Black circles show the results from this work while blue downward triangles indicate the results for the proton from the BaBar experiment~\cite{osz1}. The wide orange and fine blue dashed lines are the fit results with a common momentum frequency $C$ to the neutron and proton data with Eq.~(\ref{funcosz}). }
	\label{fig:results2}
\end{figure*}
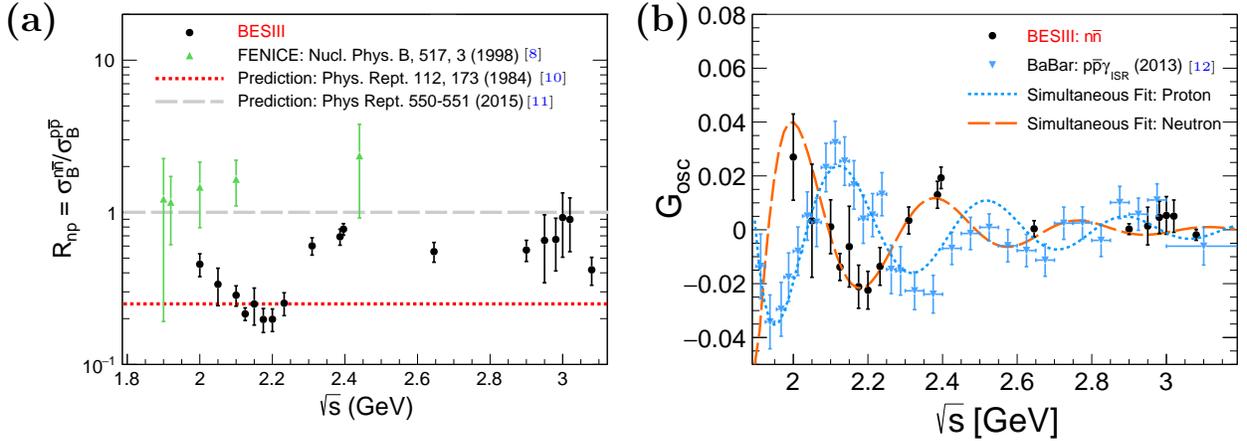

Our results significantly improve the overall precision of the available data for the neutron. For $\sqrt{s} = 2.0$, 2.1 and 2.4 GeV, the precision is improved over previous
measurements by factors of about 3, 2, and 6, respectively. They reach a comparable precision to those from the SND experiment below $\sqrt{s} = 2.0$ GeV. Our measurements are systematically below all other previously measured values above 2 GeV, while still in agreement within two standard deviations taking into account individual uncertainties. The FENICE experiment published results on the Born cross section for the process $e^+e^-\to n\bar{n}$ and  $e^+e^-\to p\bar{p}$. The average Born cross section over the center-of-mass energies yielded a ratio of $R_{np} = \sigma_B^{n\bar{n}}/\sigma_B^{p\bar{p}} = 1.69 \pm 0.49 >1$ \cite{FENICE}, indicating a stronger coupling of the virtual photon $\gamma^*(q^2)$ with the neutron than with the proton.  Using the results from this analysis and a recent publication by the BESIII experiment on the $e^+e^-\to p \bar{p}$ Born cross section~\cite{protonscan}, which was extracted from the same data samples, we test this assumption. As shown in Fig.~\ref{fig:results2}(a), our values range from 0.25 to 1 and do not support the FENICE conjecture. This result agrees with the predictions from \cite{Chernyak} and clarifies this photon-nucleon interaction puzzle that has persisted for over 20 years.

With the values for $|G|$ obtained in this analysis, an interesting feature can be similarly tested as observed for the proton results $|G_p|$ by the BaBar experiment. $|G_p|$ shows an oscillating behavior around $G_{osc} (q^2) \equiv G_{osc} $ \cite{osz1}, \cite{osz2},

\begin{linenomath*}
\begin{equation}\label{moddip}
\begin{split}
 G_{osc}(q^2) = |G| - G_{D}, \\ G_D(q^2)=\frac{\mathcal{A}_{n}}{(1-\frac{q^2}{0.71 ( \textrm{GeV}^2)})^2}
 \end{split}
\end{equation}
\end{linenomath*}
\noindent as shown in Fig. \ref{fig:results2} (right). The parameters for the normalization $\mathcal{A}_p$ and the pole $m^2_a$ have been determined from the fit to the $|G_p|$ results from the BaBar experiment as $\mathcal{A}_p = 7.7$ and $m^2_a = 14.8$ GeV$^2$. With the fixed parameter $m^2_a = 14.8$ GeV$^2$, we obtain the normalization for the neutron process $\mathcal{A}_n = 4.87 \pm 0.09$. The periodic structure $F_{osc}$ is parameterized as a function of the relative momentum $p$ which was used in the ref.~\cite{osz2}:

\begin{linenomath*}\footnotesize
\begin{equation}\label{funcosz}
\centering
\begin{split}
F_{osc}^{n,p} = A^{n,p}\cdot \textrm{exp}\left(-B^{n,p} \cdot p \right)\cdot \cos \left(C \cdot p + \it{D^{n,p}}\right),\\
p\equiv \sqrt{E^2-m^2_{n,p}},
E\equiv\frac{q^2}{2m_{n,p}}-m_{n,p}
\end{split}
\end{equation}
\end{linenomath*}
 \noindent Here, $A$ is the normalization, $B$ the inverse oscillation damping, $C$ the momentum frequency, and $D$ the phase. We perform a simultaneous fit to the neutron and proton data with a common momentum frequency $C$. The results are shown in Fig. \ref{fig:results2} (b). Our results show an almost orthogonal periodic behavior for $|G|$ of the neutron, when compared to the proton. With a common momentum frequency $C = (6.5 \pm 0.1)$ GeV$^{-1}$ and a phase difference of $\Delta D = |D_p - D_n| = (123 \pm 12)^{\circ}$ and $\chi^2/dof = 71/37$, the fit describes both data sets. Possible explanations for this oscillation are for example interference effects from final state re-scattering \cite{osz4}, or a resonant structure \cite{osz5}.

\section{Conclusion}
 We have measured the  Born cross section, $\sigma_{B}$, for the $ e^+ e^- \to  n\bar{ n}$ process and the corresponding effective form factor $|G|$ with an unprecedented precision between $\sqrt{s} = 2.0$ and 3.08 GeV. Our results are in agreement with the recent publication from the SND experiment around $\sqrt{s} = 2.0$ GeV, but not consistent with the FENICE data at higher c.m.~energies. Using recent BESIII results for $ e^+  e^-  \to  p \bar{ p}$ based on the same data set as used for our analysis, we obtained the ratio $ R_{ n p} =  \sigma_{ B}^{ n \bar{ n}}/ \sigma_{ B}^{ p \bar{ p}}$ < 1, in contradiction to the FENICE results. Our data shows that the photon-proton interaction is stronger than the  corresponding photon-neutron interaction, as expected from most theoretical predictions. The periodic structure of $|G|$ for the proton demonstrates a deviation from the dipole law. We observe a similar periodic behavior in the case of the neutron with a large phase difference $\Delta D = (125 \pm 12)^{\circ}$, when compared to the proton. Theoretical investigations as well as more experimental data could help resolve the origin for the oscillation of the electromagnetic structure observables of the nucleon. The results provide a new insight into the fundamental properties of the neutron. They can be used to constrain the parameterizations of the general parton distribution, which is closely related to the neutron spin \cite{spinf}, and are related to the neutron mass, according to the Feynman-Hellmann theorem \cite{hellmann}. Furthermore, the extracted form factors can be directly translated to the neutron radius in the Breit-Frame \cite{breit}, and can be used to understand the controversy of the neutron charge radius \cite{neutronradius1,neutronradius2}, when combined with the scattering data. The knowledge of the electromagnetic structure of the neutron is needed for the understanding of many fundamental processes. For example, the distribution of the neutron in nuclei and its structure plays a major role in the calculation of neutron star radii \cite{context1}. A possible QCD phase transition from nuclear matter to Quark-Gluon-Plasma involves neutron structure and annihilation reactions play a major role in the simulation of the measurements \cite{context2,context3}. The observation of the light curve and gravitational wave signals of a nearby neutron star merger as observed recently by gravitational wave detectors allows to identify the different contributions to this violent process in terms of nuclear physics, nucleon structure and general relativity \cite{context4}.

\bigskip
\acknowledgments
The BESIII collaboration thanks the staff of BEPCII and the IHEP computing center and the supercomputing center of USTC for their strong support. This work is supported in part by National Key R\&D Program of China under Contracts Nos. 2020YFA0406400, 2020YFA0406300; National Natural Science Foundation of China (NSFC) under Contracts Nos. 11625523, 11635010, 11735014, 11805124, 11822506, 11835012, 11935015, 11935016, 11935018, 11961141012, 12022510, 12025502, 12035009, 12035013, 12061131003, 11705192, 11950410506, 12061131003; The Chinese Academy of Sciences (CAS) Large-Scale Scientific Facility Program; Joint Large-Scale Scientific Facility Funds of the NSFC and CAS under Contracts Nos. U1732263, U1832207, U1832103, U2032111; CAS Key Research Program of Frontier Sciences under Contract No. QYZDJ-SSW-SLH040; 100 Talents Program of CAS; Guangdong Major Project of Basic and Applied Basic Research No. 2020B030103008 and Science and Technology Porgram of Guangzhou (No. 2019050001); INPAC and Shanghai Key Laboratory for Particle Physics and Cosmology; ERC under Contract No. 758462; European Union Horizon 2020 research and innovation programme under Contract No. Marie Sklodowska-Curie grant agreement No 894790; German Research Foundation DFG under Contracts Nos. 443159800, Collaborative Research Center CRC 1044, FOR 2359, GRK 214; Istituto Nazionale di Fisica Nucleare, Italy; Ministry of Development of Turkey under Contract No. DPT2006K-120470; National Science and Technology fund; Olle Engkvist Foundation under Contract No. 200-0605; STFC (United Kingdom); The Knut and Alice Wallenberg Foundation (Sweden) under Contract No. 2016.0157; The Royal Society, UK under Contracts Nos. DH140054, DH160214; The Swedish Research Council; U. S. Department of Energy under Contracts Nos. DE-FG02-05ER41374, DE-SC-0012069.\\

\appendix{ \section{}}
\subsection{ {Results for category A, B, C} \label{app:results}}
\begin{figure*}[htb]
	\centering
\includegraphics[width=0.48\textwidth]{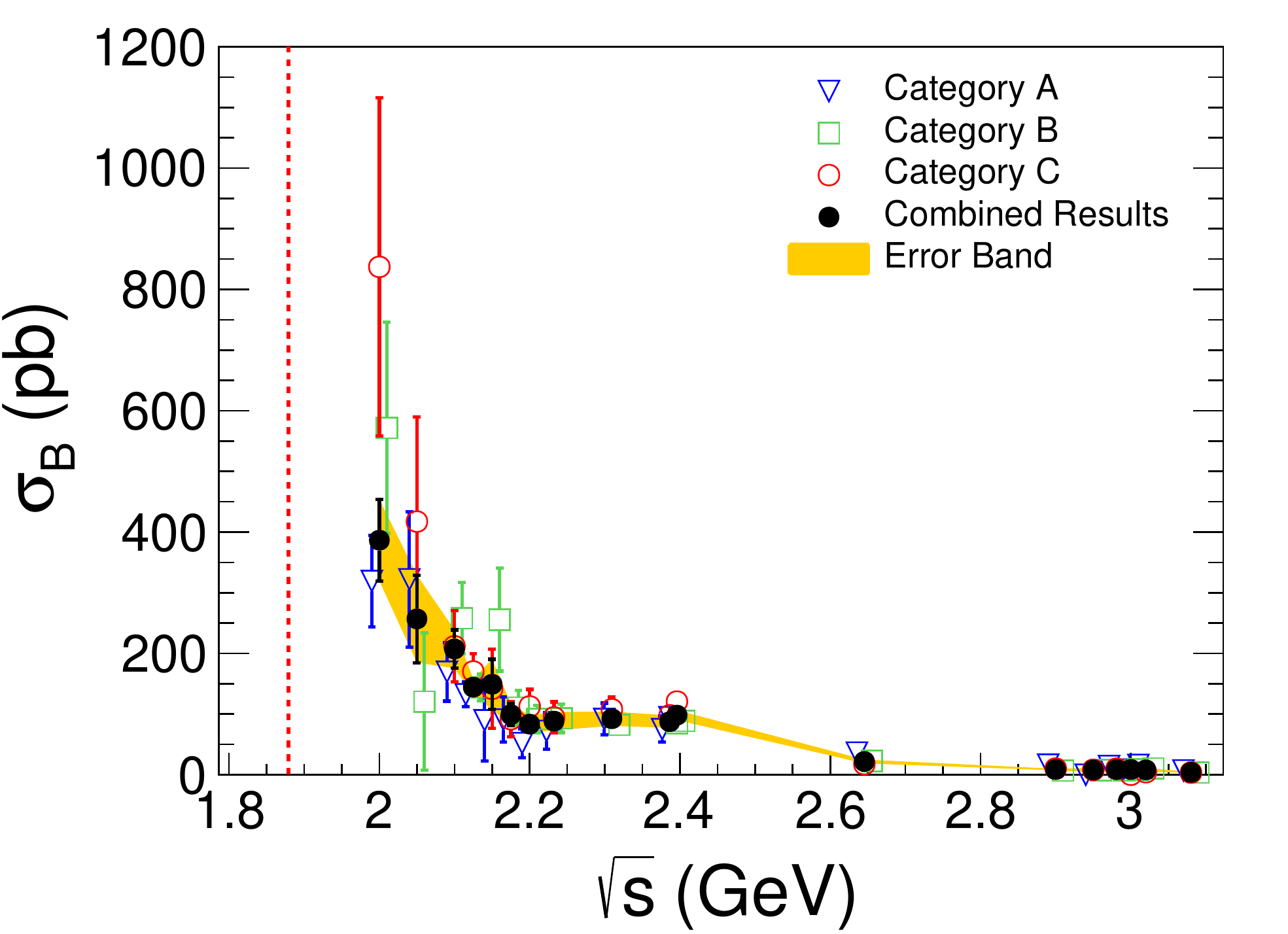}
\includegraphics[width=0.48\textwidth]{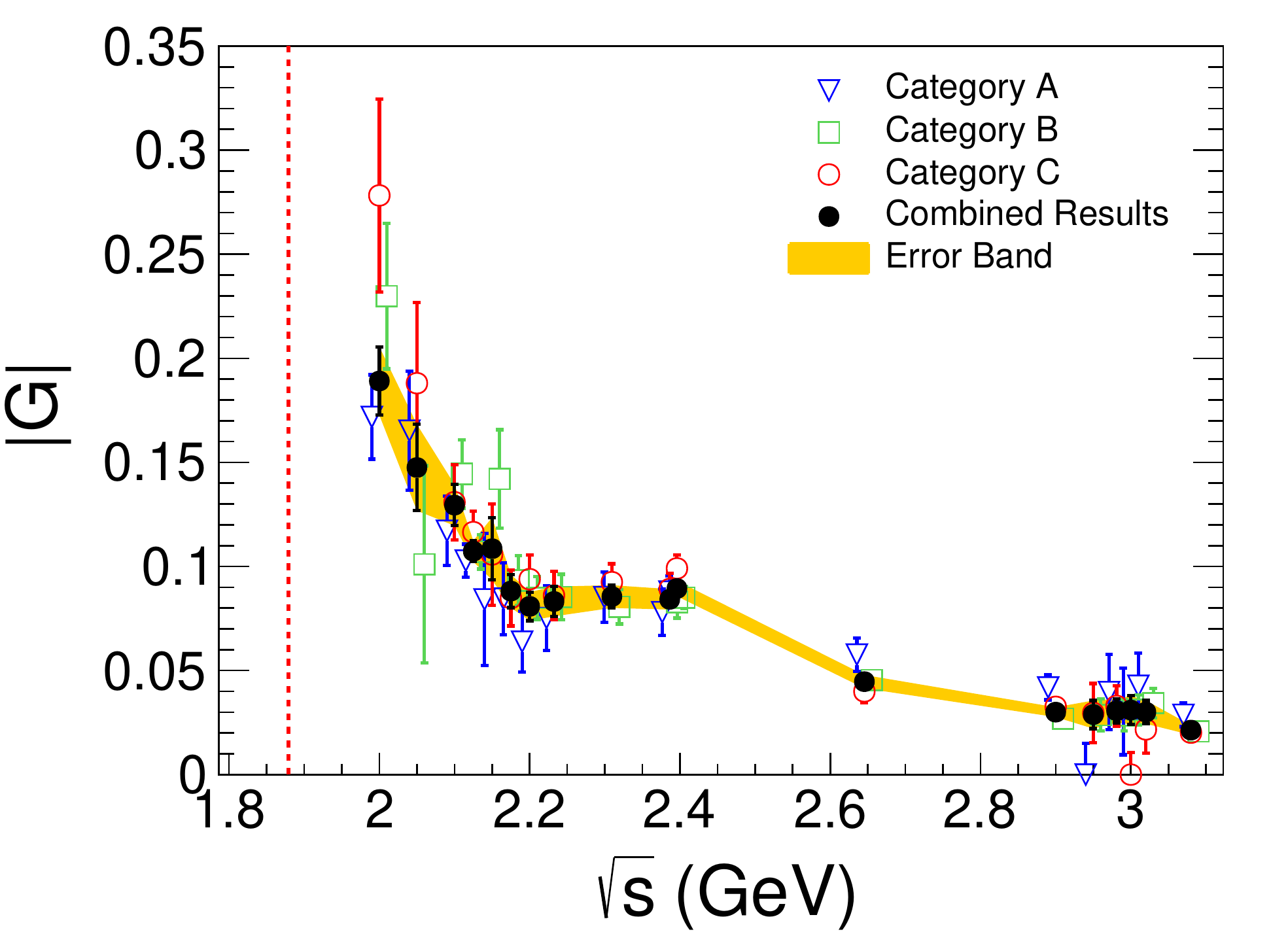}
\caption[Comparison of the results for the Born cross section and the effective form factor for the process $e^+e^-\to n\bar{n}$ from the three signal classification categories.]{Comparison of the results for (left) the Born cross section and (right) the effective form factor for the process $e^+e^-\to n\bar{n}$ from the three signal classification categories. Results shown in blue, green and red are extracted with the signal classification category A, B, and C, respectively, all with statistical uncertainty only. The black open circles are the combined results using the formula from Eq.~(\ref{eqn:weightedXS1}). The yellow bands represents the statistical uncertainty of the combined results as listed in Table \ref{tab:results} of the main text. The individual results are listed in the Tables \ref{tab:XSandEFF} - \ref{tab:XSandEFF3}.
}
\label{c9:fig:bornsum}
\end{figure*}

Figure~\ref{c9:fig:bornsum} illustrates
 a detailed comparison of the results for the Born cross section and the effective form factor for the process $e^+e^-\to n\bar{n}$ from the three signal classification categories.
 The individual results are listed in the Tables \ref{tab:XSandEFF} - \ref{tab:XSandEFF3}.

\begin{table*}[htb]\footnotesize
 \tabcolsep=0.095cm
  \renewcommand{\arraystretch}{1.2}
\begin{center}
\caption{Summary of the results $\sigma_{B}^{A}$, $|G^{A}|$, and the related quantities at each $\sqrt{s}$. The first uncertainty is statistical and the second one systematic.}
\begin{tabular}{lcccccccc}
  \hline
  \hline
  $\sqrt{s}$ (GeV) &  $N^{s}_A$  &  $\mathcal{L}_{\rm int}$~(pb$^{-1}$) & $(1+\delta)_A$  & $\varepsilon_{MC}^A(\%)$ & $\mathcal{C}_{n\bar{n}}(\%)$  &  $\mathcal{C}_{trg}(\%)$ & $\sigma_{B}^A$ (pb) & $|G^A|$ $(\times 10^{-2})$  \\
  \hline
  2.0000    &$  38.3\pm 7.3$  &10.1  & 0.98 &  1.35   & 108.3  & 82.9  & $  319\pm60  \pm44  $ &   $17.3\pm1.6\pm1.2$\\
  2.0500    &$  12.8\pm 4.1$  &3.34  & 1.08 &  1.27   & 104.1  & 83.3  & $  320\pm100 \pm40  $ &   $16.5\pm2.6\pm1.0$\\
  2.1000    &$  24.3\pm 5.7$  &12.2  & 1.18 &  1.23   &  96.5  & 83.8  & $  169\pm40  \pm28  $ &   $11.7\pm1.4\pm1.0$\\
  2.1250    &$   172\pm15$    &108   & 1.24 &  1.19   &  96.3  & 84.6  & $  133\pm11  \pm17  $ &   $10.3\pm0.4\pm0.7$\\
  2.1500    &$   3.0\pm 2.2$  &2.84  & 1.29 &  1.15   &  93.3  & 85.7  & $ 89\pm65\pm14$ &   $ 8.4\pm3.1\pm0.7$\\
  2.1750    &$  10.9\pm 4.1$  &10.6  & 1.31 &  1.09   &  94.4  & 83.8  & $ 91\pm34\pm14$ &   $ 8.5\pm1.6\pm0.7$\\
  2.2000    &$   8.0\pm 3.5$  &13.7  & 1.31 &  1.06   &  94.6  & 85.3  & $ 52\pm22\pm7 $ &   $ 6.4\pm1.4\pm0.4$\\
  2.2324    &$  10.0\pm 3.9$  &11.9  & 1.28 &  1.08   &  98.8  & 85.1  & $ 72\pm28\pm10$ &   $ 7.5\pm1.5\pm0.5$\\
  2.3094    &$  22.6\pm 5.8$  &21.1  & 1.14 &  1.21   &  96.6  & 87.6  & $ 91\pm23\pm11$ &   $ 8.5\pm1.1\pm0.5$\\
  2.3864    &$  22.5\pm 5.8$  &22.5  & 1.11 &  1.40   &  95.9  & 89.3  & $ 75\pm19\pm9 $ &   $ 7.8\pm1.0\pm0.5$\\
  2.3960    &$  80.3\pm 9.9$  &66.9  & 1.11 &  1.35   &  94.5  & 89.2  & $ 95\pm11\pm11$ &   $ 8.8\pm0.5\pm0.5$\\
  2.6454    &$  19.4\pm 4.7$  &67.7  & 1.55 &  0.64   &  83.8  & 94.7  & $ 37\pm9 \pm5 $ &   $ 5.8\pm0.7\pm0.4$\\
  2.9000    &$  16.3\pm 4.4$  &105   & 2.16 &  0.56   &  79.9  & 95.9  & $ 17\pm5 \pm2 $ &   $ 4.2\pm0.6\pm0.2$\\
  2.9500    &$   0.0\pm 1.3$  &15.9  & 2.29 &  0.57   &  77.9  & 96.7  & $  0.0\pm8.3 \pm0.0 $ &   $        -       $\\
  2.9810    &$   2.3\pm 1.9$  &16.1  & 2.36 &  0.56   &  78.8  & 95.8  & $ 15\pm11\pm5 $ &   $ 4.1\pm1.5\pm0.7$\\
  3.0000    &$   1.4\pm 1.3$  &15.9  & 2.41 &  0.57   &  80.2  & 96.0  & $  8.3\pm7.7 \pm8.4 $ &   $ 3.0\pm1.4\pm1.5$\\
  3.0200    &$   2.9\pm 2.1$  &17.3  & 2.46 &  0.55   &  80.0  & 95.7  & $ 16\pm11\pm3 $ &   $ 4.2\pm1.5\pm0.4$\\
  3.0800    &$  12.1\pm 4.3$  &126   & 2.61 &  0.56   &  95.8  & 96.5  & $  7.1\pm2.5 \pm1.3 $ &   $ 2.9\pm0.5\pm0.3$\\
\hline
\hline
\end{tabular}
\label{tab:XSandEFF}
\end{center}
\end{table*}

\begin{table*}[htb]\footnotesize
 \tabcolsep=0.095cm
  \renewcommand{\arraystretch}{1.2}
\begin{center}
\caption{Summary of the results $\sigma_{B}^{B}$, $|G^{B}|$, and the related quantities at each $\sqrt{s}$. The first uncertainty is statistical and the second one systematic.}
\begin{tabular}{lcccccccc}
  \hline
  \hline
  $\sqrt{s}$ (GeV) &  $N^{s}_B$  &  $\mathcal{L}_{\rm int}$~(pb$^{-1}$) & $(1+\delta)_B$  & $\varepsilon_{MC}^B(\%)$ & $\mathcal{C}_{n\bar{n}}(\%)$  &  $\mathcal{C}_{trg}(\%)$ & $\sigma_{B}^B$ (pb) & $|G^B|$ $(\times 10^{-2})$  \\
  \hline
  2.0000 &$ 16.2 \pm4.5  $  &10.1 & 0.98 & 0.43 & 76.8  & 86.7   &$  570\pm160 \pm70  $&  $ 23.1\pm3.2\pm1.4$\\
  2.0500 &$ 2.27 \pm2.1  $  &3.34 & 1.08 & 0.86 & 69.6  & 87.2   &$  120\pm110 \pm20  $&  $ 10.1\pm4.6\pm0.8$\\
  2.1000 &$ 36.1 \pm6.6  $  &12.2 & 1.18 & 1.43 & 76.6  & 88.7   &$  258\pm47  \pm35  $&  $ 14.5\pm1.3\pm1.0$\\
  2.1250 &$ 226  \pm17   $  &108  & 1.24 & 1.69 & 77.8  & 89.4   &$  144\pm11  \pm19  $&  $ 10.7\pm0.4\pm0.7$\\
  2.1500 &$ 12.5 \pm3.8  $  &2.84 & 1.28 & 1.91 & 78.0  & 90.3   &$  256\pm78  \pm34  $&  $ 14.2\pm2.2\pm0.9$\\
  2.1750 &$ 22.7 \pm5.2  $  &10.6 & 1.31 & 2.04 & 81.7  & 88.6   &$  110\pm25  \pm12  $&  $  9.3\pm1.1\pm0.5$\\
  2.2000 &$ 29   \pm5.8  $  &13.7 & 1.31 & 2.38 & 81.6  & 90.4   &$   92\pm19\pm12$&  $  8.5\pm0.9\pm0.6$\\
  2.2320 &$ 27.3 \pm5.9  $  &11.9 & 1.28 & 2.56 & 83.1  & 90.5   &$   93\pm20\pm12$&  $  8.5\pm0.9\pm0.6$\\
  2.3090 &$ 40.8 \pm6.8  $  &21.1 & 1.15 & 2.56 & 85.9  & 94.0   &$   82\pm14\pm10 $&  $  8.1\pm0.7\pm0.5$\\
  2.3860 &$ 56.9 \pm8.0  $  &22.5 & 1.11 & 3.20 & 88.7  & 94.8   &$   85\pm12\pm10$&  $  8.3\pm0.6\pm0.5$\\
  2.3960 &$ 172  \pm14   $  &66.9 & 1.11 & 3.12 & 88.7  & 94.3   &$   89\pm7 \pm8 $&  $  8.5\pm0.3\pm0.4$\\
  2.6440 &$ 57.9 \pm8.2  $  &67.7 & 1.55 & 3.02 & 84.0  & 96.3   &$   22\pm3 \pm3 $&  $  4.5\pm0.3\pm0.3$\\
  2.9000 &$ 29.3 \pm5.8  $  &105  & 2.17 & 2.40 & 80.7  & 97.2   &$  6.9\pm1.4 \pm0.9 $&  $  2.7\pm0.3\pm0.2$\\
  2.9500 &$ 5.1  \pm2.6  $  &15.9 & 2.29 & 2.34 & 80.7  & 97.7   &$  7.6\pm3.9 \pm1.1 $&  $  2.9\pm0.7\pm0.2$\\
  2.9810 &$ 4.92 \pm2.5  $  &16.1 & 2.36 & 2.22 & 80.7  & 97.3   &$  7.4\pm3.8 \pm1.1 $&  $  2.9\pm0.7\pm0.2$\\
  3.0000 &$ 5.87 \pm2.6  $  &15.9 & 2.41 & 2.24 & 80.7  & 97.5   &$  8.7\pm3.8 \pm1.4 $&  $  3.1\pm0.7\pm0.3$\\
  3.0200 &$ 7.51 \pm2.8  $  &17.3 & 2.46 & 2.14 & 80.7  & 97.4   &$   10\pm4 \pm2 $&  $  3.4\pm0.5\pm0.3$\\
  3.0800 &$ 19.8 \pm4.9  $  &126  & 2.61 & 2.08 & 80.0  & 97.8   &$  3.7\pm0.9 \pm0.7 $&  $  2.1\pm0.3\pm0.2$\\

\hline
\hline
\end{tabular}
\label{tab:XSandEFF2}
\end{center}
 \tabcolsep=0.095cm
 \renewcommand{\arraystretch}{1.2}
\begin{center}
\caption{Summary of the results $\sigma_{B}^{C}$, $|G^{C}|$, and the related quantities at each $\sqrt{s}$. The first uncertainty is statistical and the second one systematic.}
\begin{tabular}{lccccccccc}
  \hline
  \hline
  $\sqrt{s}$ (GeV) &  $N^{s}_C$  &  $\mathcal{L}_{\rm int}$~(pb$^{-1}$) & $(1+\delta)_C$  & $\varepsilon_{MC}^C(\%)$ & $\mathcal{C}_{\rm{n\bar{n}}}(\%)$ & $\mathcal{C}_{\textrm{trg}}(\%)$ &  $\mathcal{C}_{\textrm{ee}}(\%)$ & $\sigma_{B}^C$ (pb) & $|G^C|$ $(\times 10^{-2})$  \\
  \hline
  2.0000     &$  25.8 \pm 6.0 $ &  10.1  & 0.98 & 0.34 & 137.5  & 88.8 &  75.0 &$  840\pm200  \pm200 $ &  $28.0\pm  3.3\pm  3.3$\\
  2.0500     &$  9.8  \pm 3.5 $ &  3.34  & 1.08 & 0.75 & 126.7  & 91.1 &  75.2 &$  420\pm150  \pm90  $ &  $18.9\pm  3.4\pm  2.0$\\
  2.1000     &$  31.3 \pm 6.2 $ &  12.2  & 1.18 & 1.15 & 125.4  & 92.1 &  77.3 &$  212\pm42   \pm41  $ &  $13.1\pm  1.3\pm  1.3$\\
  2.1266     &$ 281   \pm 18  $ &  108   & 1.24 & 1.34 & 124.7  & 92.5 &  79.8 &$  170\pm11   \pm27  $ &  $11.7\pm  0.4\pm  0.9$\\
  2.1500     &$  7.4  \pm 3.1 $ &  2.84  & 1.28 & 1.51 & 123.5  & 92.9 &  83.0 &$  142\pm59   \pm27  $ &  $10.6\pm  2.2\pm  1.0$\\
  2.1750     &$ 19.7  \pm 5.4 $ &  10.6  & 1.31 & 1.59 & 123.4  & 91.8 &  86.0 &$   92\pm25   \pm14  $ &  $ 8.5\pm  1.2\pm  0.7$\\
  2.2000     &$ 35.2  \pm 6.9 $ &  13.7  & 1.31 & 1.74 & 123.1  & 92.3 &  88.1 &$  113\pm22   \pm17  $ &  $ 9.4\pm  0.9\pm  0.7$\\
  2.2324     &$  27.4 \pm 6.0 $ &  11.9  & 1.28 & 1.89 & 121.6  & 92.9 &  89.2 &$   94\pm20 \pm15$ &  $ 8.6\pm  0.9\pm  0.7$\\
  2.3094     &$  52.3 \pm 8.1 $ &  21.1  & 1.14 & 1.93 & 122.7  & 95.0 &  89.6 &$  108\pm17 \pm12    $ &  $ 9.3\pm  0.7\pm  0.5$\\
  2.3864     &$  57.0 \pm 8.0 $ &  22.5  & 1.11 & 2.32 & 116.7  & 95.9 &  89.7 &$   98\pm14 \pm9 $ &  $ 8.9\pm  0.6\pm  0.4$\\
  2.3960     &$ 212   \pm 16  $ &  66.9  & 1.11 & 2.36 & 116.7  & 95.8 &  89.7 &$  121\pm9  \pm13$     &  $ 9.9\pm  0.4\pm  0.5$\\
  2.6454     &$  44.1 \pm 8.6 $ &  67.7  & 1.55 & 2.74 & 100.6  & 97.1 &  89.8 &$   17\pm3  \pm4 $ &  $ 4.1\pm  0.3\pm  0.5$\\
  2.9000     &$  47.5 \pm 8.5 $ &  105   & 2.16 & 2.28 & 103.2  & 97.6 &  90.0 &$   10\pm2  \pm2 $ &  $ 3.2\pm  0.3\pm  0.3$\\
  2.9500     &$  6.1  \pm 3.3 $ &  15.9  & 2.29 & 2.25 & 104.6  & 98.0 &  90.1 &$  8.1\pm4.4  \pm6.4 $ &  $ 3.0\pm  0.8\pm  1.2$\\
  2.9810     &$  7.5  \pm 3.8 $ &  16.1  & 2.36 & 2.16 & 104.8  & 97.7 &  90.1 &$  9.9\pm5.0  \pm3.0 $ &  $ 3.3\pm  0.8\pm  0.5$\\
  3.0000     &$    0  \pm 3.0 $ &  15.9  & 2.41 & 2.11 & 105.6  & 97.7 &  90.1 &$  0.0\pm4.0  \pm0.0 $ &  $           -        $    \\
  3.0200     &$  3.4  \pm 3.4 $ &  17.3  & 2.46 & 2.06 & 105.7  & 97.8 &  90.1 &$  4.2\pm4.2  \pm1.4 $ &  $ 2.2\pm  1.1\pm  0.4$\\
  3.0800     &$  21.0 \pm 6.2 $ &  126   & 2.61 & 1.93 & 105.9  & 97.9 &  90.2 &$  3.5\pm1.0  \pm1.0 $ &  $ 2.0\pm  0.3\pm  0.3$\\
\hline
\hline
\end{tabular}
\label{tab:XSandEFF3}
\end{center}
\end{table*}

\subsection{Monte Carlo Simulation \label{app:mc}}{

\noindent{\bf Monte Carlo (MC) simulations used in this analysis}. Signal Monte Carlo samples have been produced for the optimization of the signal selection, the determination of the signal efficiency, and the estimation of the corrections from the QED Initial-State-Radiation (ISR) events. The signal MC sample is produced with the generator {\sc conexc} \cite{signalMC} which is designed to simulate events up to the Next-to-Leading Order (NLO) and using the implementation of the vacuum polarization by Jegerlehner \cite{jeger}. Since the $n\bar{n}$ final state is electrically neutral, no QED Final-State-Radiation (FSR) effects occur. Background from multi-hadronic processes is estimated from MC simulations generated with {\sc lund} \cite{LUND}. Cross sections and angular distributions from measured processes are implemented in the generator, while unmeasured processes are generated as phase-space. Background from QED processes, such as $e^+e^- \rightarrow e^+e^-$ or $e^+e^-\rightarrow\gamma\gamma$ are generated with {\sc babayaga} \cite{babayaga} in Next-to-Next-Leading-Order (NNLO) including vacuum polarization, ISR- and FSR-effects. Finally, we use MC simulation for the control channels $e^+e^- \rightarrow J/\psi \rightarrow p\bar{n}\pi^- (\bar{p}n\pi^+)$, which are generated with {\sc KKMC}~\cite{KKMC}. All MC simulation are generated according to the integrated luminosity of the collider data, containing equal or larger numbers of events then available from data. \\

\subsection{ { \textbf Methods} } \label{app:rec}

\noindent{\bf TOF based algorithm for the reconstruction of neutral particles}. A common method in signal classification category A and B, which combines the response from the EMC and the TOF, is described as follows. The most energetic shower in an event is identified as the anti-neutron $\bar{n}$. Its position vector in the EMC response ${\bf V_{EMC1}}$ with respect to the $e^+e^-$ interaction point (IP) ${\bf V_{IP}}$ is associated with the closest TOF response with the position vector ${\bf V_{TOF1}}$, if the distance in the TOF plane $\Delta^{EMC1}_{TOF1} = |{\bf V_{EMC1}} - {\bf V_{TOF1}}|$ is smaller than the azimuthal span of 3 TOF counters. The flight length of the anti-neutron to the TOF response is $L_{\bar{n}} = |{\bf V_{TOF1}} - {\bf V_{IP}}|$. The flight time of the anti-neutron $T_{TOF1}$ is determined by an algorithm \cite{est} using the hypothesis of a photon producing the TOF response. The expected flight time for a photon from the IP to the TOF response ${\bf V_{TOF1}}$ is $T_{\gamma}^{exp} = L_{\bar{n}}/c$, where $c$ is the speed of light in vacuum. For the time difference $\Delta T_{\bar{n}} = T_{TOF1} - T_{\gamma}^{exp}$, values different from zero are expected for anti-neutron, therefore this criterion can be used for the discrimination against photon background. A similar approach is chosen for the reconstruction of the neutron candidate $n$.  The time difference $\Delta T_n = T^{obs}_{n} - T_n^{exp}$ is used to identify $n$ candidates. $T_n^{obs}$ is the measured time, $T_n^{exp} = L_{n}/(\beta c)$ the expected flight time under the hypothesis of a neutron, where $L_{n}$ is the flight length of the neutron to the coordinates of the TOF response. Furthermore, the opening angle between the anti-neutron position vector ${\bf V_{EMC1}}$ and the measured TOF position vector ${\bf V_{TOF2'}}$ can be used to suppress events with more than two final state particles, beam-associated background. The process of $e^+e^-\to J/\psi\to \pi^+\pi^-\pi^0$ is used to verify the photon detection efficiency with above method. Fig. \ref{fig:photondetectioneffiency} shows that the data is in an excellent agreement with MC simulation (difference $<1\%$). Additionally, we verify the efficiency of the neutral TOF reconstruction with the well known channel $e^+e^-\to \gamma\gamma$, as well as with $e^+e^-\to J/\psi \to n\bar{n}$. The results for the cross section and branching fraction are in excellent agreement with the world reference, as shown in the Tables \ref{tab:digamma} and \ref{tab:jpsi2nnbar}, respectively.\\

\noindent Additionally,
 the process $e^+e^-\to\gamma\gamma$ is used to verify the TOF-based algorithm for the time-of-flight reconstruction of neutral particles. The selection criteria are as follows: the event must contain no charged tracks and at least one showers in the EMC, the most energetic shower must be within $|\cos\theta | < 0.8$, the difference between the measured and expected flight time for the most energetic EMC shower and the second TOF hit must be within $|T^{\gamma}_{TOFi}-T_{exp}^{\gamma}|<1$ ns ($i=1,2$), the deposition energy of the most energetic EMC shower $E_1$ must be within $0.7 \sqrt{s}/2 < E_1 < 1.1 \sqrt{s}/2$ (GeV).
The time difference between the measured time of two showers must be within $|T_{TOF1}^{\gamma}-T_{TOF2}^{\gamma}|<1$ ns considering the identical features of two photons.
The open angle between the position vector of the leading energetic shower and the measured position vector of the second TOF  must be larger than 3.00 radian.
The results for the cross section and branching fraction are in excellent agreement with the world reference, as shown in the Table~\ref{tab:digamma} and the Table~\ref{tab:jpsi2nnbar} for the process $e^+e^-\to J/\psi \to n\bar{n}$ in the \textbf{Appendix}~\ref{app:crosscheck}.  \\
\begin{figure}[htb]
  \centering
  \includegraphics[width=7.5cm]{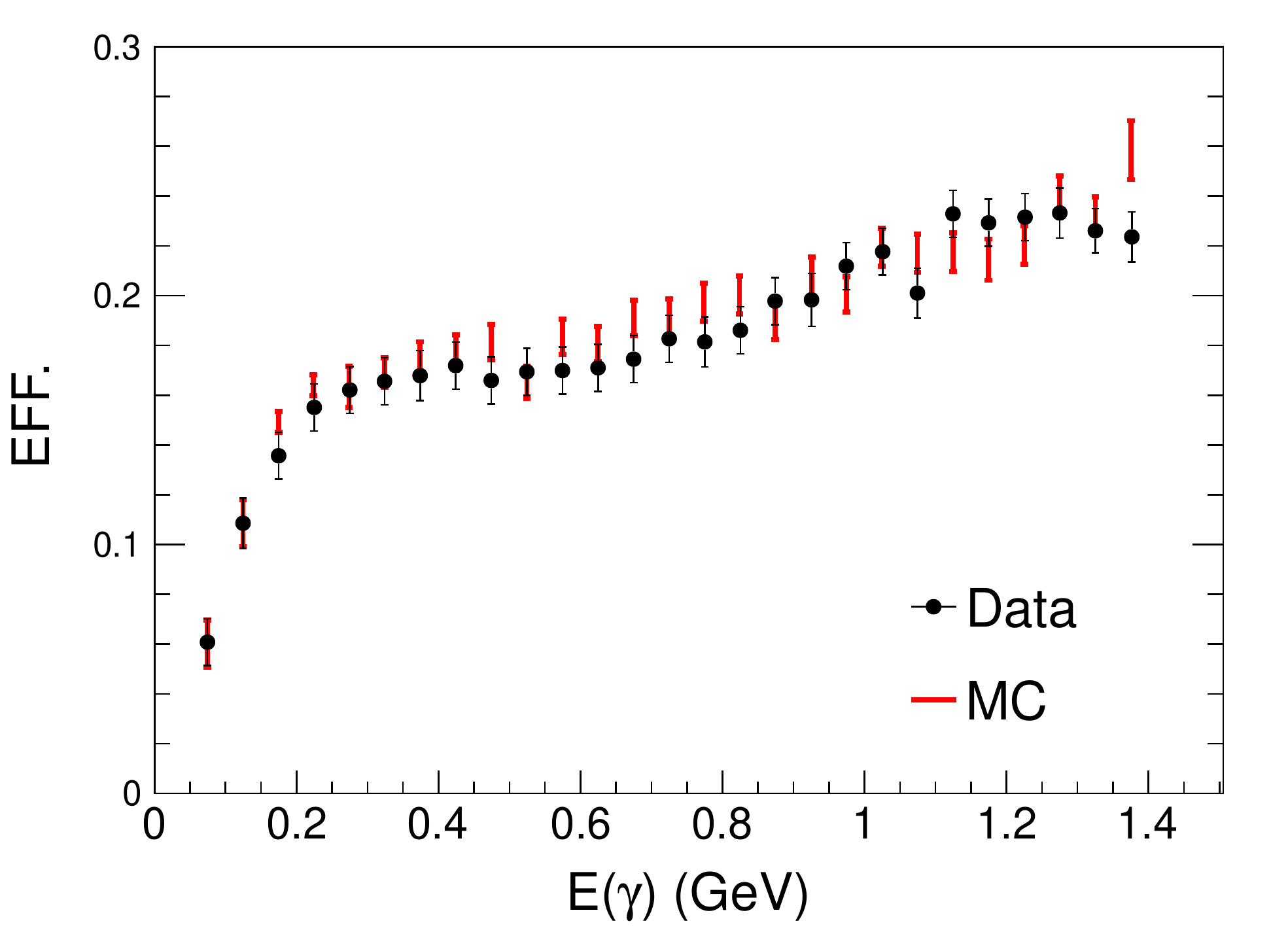}
  \includegraphics[width=7.5cm]{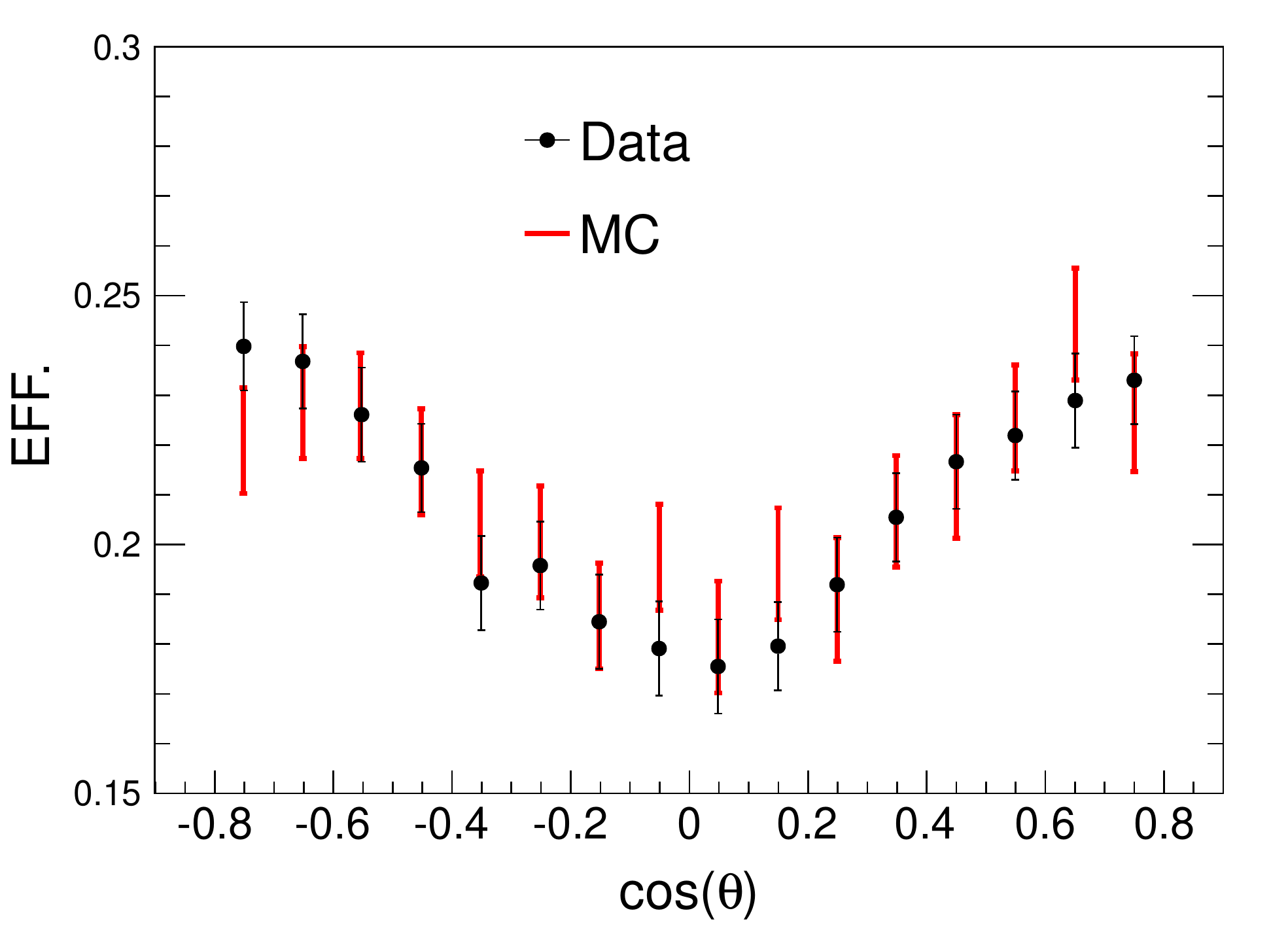}
    \caption{Reconstruction efficiency for a photon using the neutral TOF algorithm (left) as a function of the deposition energy in the EMC with $\cos\theta$ values within ($-$0.8, 0.8), and (right) as a function of $\cos\theta$ with values for the deposition energy within (0.80, 1.20) GeV.}\label{fig:photondetectioneffiency}
\end{figure}

 \begin{table}[htb]\footnotesize
 \begin{center}
   \caption{Verification of the reconstruction efficiency for the neutral TOF algorithm with the process $e^+e^-\to\gamma\gamma$. The luminosity is quoted from \cite{luminosity}, the reference Born cross section $\sigma_{ref}$ is taken from {\sc babayaga} generator~\cite{babayaga}, the reconstruction efficiency $\varepsilon_{MC}^{\gamma\gamma}$ is determined from dedicated signal MC samples generated with \cite{babayaga}, the observed cross section is determined as $\sigma_{obs}=N_{\gamma\gamma}/(\mathcal{L}\varepsilon_{MC}^{\gamma\gamma})$. The uncertainties are statistical only.}
 \begin{tabular}{cccccc}
   \hline
   \hline
   $\sqrt{s}$ (GeV) & $\mathcal{L}$ (pb$^{-1}$) & $\varepsilon_{MC}^{\gamma\gamma}$(\%) & $N_{\gamma\gamma}$ & $\sigma_{obs}$ (nb) & $\sigma_{ref}$ (nb) \\
   \hline
   \hline
   2.0000            & $10.1$      & 2.0    & 17808   & $87.6\pm0.7$ & $88.0\pm0.9$ \\
   2.1250            & $108$    & 2.0    & 167428  & $76.2\pm0.2$ & $77.4\pm0.3$\\
   2.1750            & $10.6$      & 2.0    & 15737   & $72.9\pm0.6$ & $74.6\pm0.3$\\
   2.2000            & $13.7$      & 2.0    & 19867   & $71.2\pm0.5$ & $72.5\pm0.3$\\
   2.2324           & $11.9$      & 2.0    & 16827   & $69.7\pm0.5$ & $70.5\pm0.3$\\
   2.3960            & $66.9$      & 2.0    & 82185   & $60.2\pm0.2$ & $61.1\pm0.3$\\
   2.6444           & $33.7$     & 2.1    & 67623   & $48.4\pm0.2$ & $50.0\pm0.2$\\
   2.9000            & $105$    & 2.1    & 89826   & $41.2\pm0.1$ & $41.7\pm0.2$\\
   3.0800            & $126$    & 2.1    & 95549   & $36.8\pm0.1$ & $37.2\pm0.2$\\
   \hline
   \hline
 \end{tabular}
 \label{tab:digamma}
 \end{center}
 \end{table}

\noindent{\bf Cosmic ray background rejection with the Muon Counter}. We use the Muon counter system (MUC) to reject cosmic ray background. With the 9 layers of resistive plates (in barrel region) with iron absorbers in between, the developed algorithm is capable of distinguishing between the impact from particles coming from the IP of the BESIII detector and the response from the cosmic ray particles entering the detector from the outside. We evaluate the available information from the MUC and apply a criterion on the last layer with a hit response. A detailed validation of the efficiency for this method has been performed by studying the MUC response for neutrons, anti-neutrons and photons from the dedicated MC samples and collision data using the processes $e^+e^- \to J/\psi \to p\bar{n}\pi^-$, $e^+e^- \to J/\psi \to \bar{p}n\pi^+$, and $e^+e^- \to \gamma\gamma$, respectively. The detector response from the cosmic ray background is studied from the non-collision data samples at $\sqrt{s} = 2.2324$ GeV and $\sqrt{s} = 2.6444$ GeV and is stable over the analysed energy range. The data and MC simulation are in agreement, as shown in an example with collision data, non-collision data and MC simulation for dedicated signal background processes in the Fig. \ref{fig:muc} for the category C signal samples at $\sqrt{s} = 2.396$ GeV.\\

\begin{figure}[htb]
  \centering
  \includegraphics[width=7.5cm]{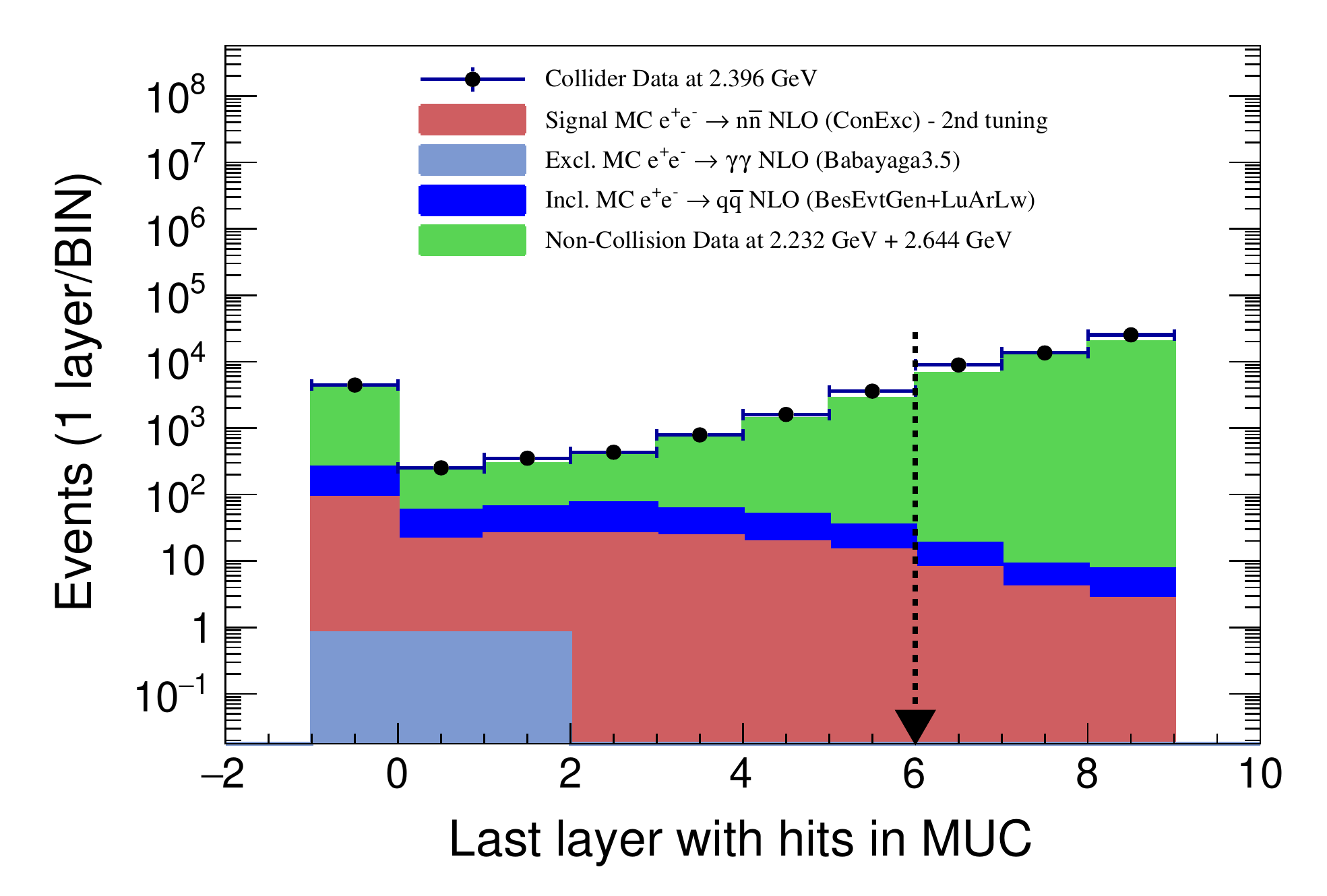}\\
  \caption{The last layer with a response among the resistive plates of the MUC for the signal reconstruction in category C at $\sqrt{s} = 2.396$ GeV. Black dots with error bars represent the distribution from collider data, the histogram in red is the signal MC simulation for $e^+e^- \to n \bar{n}$, the light blue histogram is MC simulation for $\rm e^+e^-\to \gamma\gamma$ background. The dark blue histogram represents multi-hadronic background, the green histogram is the distribution from beam-associated background using combined samples at 2.2324 and 2.6444 GeV. The black arrow indicates the selection criterion used for categories B and C.}\label{fig:muc}
\end{figure}

\subsection{ {Selections} }\label{app:sel}
\noindent{\bf General signal event reconstruction criteria}. Signal events must contain no charged tracks, one or two reconstructed showers in the EMC as the neutron $n$ and anti-neutron $\bar{n}$ candidates. The most energetic shower ($\bar{n}$) must be within $|\cos\theta | < 0.8$ and have a minimum energy deposition of 0.5 GeV.\\

\noindent{\bf Signal classification category A}. Signal events must satisfy the following requirements for the anti-neutron selection: The number of hits in the EMC in a 50 degree cone around the anti-neutron shower $N^{50}_{hits}$ must be within the window of $30 < N^{50}_{hits} <140$. Hits in the EMC are defined as signal responses from particle showers. The reconstructed position of a EMC hit is the center of gravity from the shower. The most energetic shower must be within $|\cos\theta | < 0.7$ to ensure an efficient rejection of the process $e^+e^- \to \gamma \gamma $. To select the neutron shower, a cut on $| \Delta T_n | < 4$ ns is applied. If a second EMC shower is found, a requirement of $ 0.06 < E_n < 0.70$ GeV on the deposited energy is applied at $\sqrt{s}\geq 2.6444$ GeV. The different energy cuts are performed because high momentum neutrons easily penetrate into the EMC, and thus an extra energy cut is applied to improve the signal-background ratio without reducing signal efficiency. To further suppress physics and beam-associated background, we anticipate the back-to-back kinematics of the signal process. We require the the opening angles between ${\bf V_{EMC1}}$ and ${\bf V_{TOF2'}}$ and between ${\bf V_{EMC1}}$ and ${\bf V_{EMC2}}$ to be larger than 3 radian. The flight time difference between the two final state particles $\Delta T_{n\bar{n}} = |T_{TOF1} - T_{TOF2}|$ is required to be smaller than 4 ns.\\

\noindent{\bf Signal classification category B}. To select an anti-neutron, we require $|\Delta T_{\bar{n}}| > 0.5$ ns. The energy deposition from the EMC shower associated with the anti-neutron must be within $0.5 < E_{\bar{n}} <2.0$ GeV. The neutron energy deposition in the EMC must be within $0.06 < E_n <0.60$ GeV. Both particles must be reconstructed within $|\cos\theta | < 0.75$. No signal in the MUC from the last three layers is allowed, which ensures a good rejection power against cosmic ray background. Finally, a Boosted Decision Tree (BDT) method is used to significantly reduce the remaining background. The BDT uses multiple observables from the EMC and TOF systems with a discriminator requirement of $>0.1$ for signal event reconstruction.\\

\noindent{\bf Signal classification category C}. An anti-neutron is reconstructed by requiring for the energy deposition $0.5 < E_{\bar{n}} < 2.0$ GeV, for the EMC shower position $|\cos\theta | < 0.75$, for the second moment of the anti-neutron shower $\sum E_i r_i^2/E_i >20$ cm$^2$ with $E_i$ the deposition energy in the $i$-th crystal and $r_i$ the distance between the center of the $i$-th crystal and the center of gravity of the shower. The number of hits in the EMC within of 50 degrees around the $\bar{n}$ position is required to be $35 < N^{50}_{hits} <100 $. The neutron reconstruction is the same as for category B classified signal events. To further suppress the remaining background, the same requirement as in category B is applied on the MUC, the opening angle between the EMC showers from $n$ and $\bar{n}$ must be larger than 150 degrees and the total energy deposition $E_{extra}$ outside of 50 degrees cones around the neutron and anti-neutron shower position in the EMC must be smaller than 0.15 GeV. \\

\subsection{Fitting}\label{app:sig}
\noindent{\bf Determination of the number of the signal events}. To determine the number of reconstructed signal events $\mathcal{N}_i^s$, a composite model fit $\mathcal{F}_i$ is performed to the distribution of $\Delta T_n$ for category A events and to $\sphericalangle_{n}^{\bar{n}}$ for category B and C events, as discussed in the main part of the paper. The background normalizations are determined using the luminosity of the data samples and the theoretical cross sections for the contributing processes.  The normalization for the beam-associated background is obtained using the data taking time of the non-collision and the collision samples when applicable, or via curve fitting of the background event distribution.  The fit optimization for each category is performed by minimizing the global negative log-Likelihood (NLL) with the {\sc MIGRAD} \cite{midgard} package by means of a modified version of the David-Fletcher-Powell method \cite{fletcher} taking into account the 18 local NLLs from each data set. A {\sc HESSE} \cite{midgard} algorithm calculates a full second-derivatives matrix of the model parameters space to improve the uncertainty determination. The following MINOS error analysis is performed for a further optimization of the parameter errors estimation. While the globally optimized solution may be not optimal at a specific $\sqrt{s}$, this approach improves the fit stability. The optimized fit for the three signal classification categories A, B, and C are shown for the data at $\sqrt{s} = 2.3960$ GeV in the Fig. \ref{fig:extraction}.\\

\subsection{Efficiency Corrections \label{app:eff}}
\noindent{\bf The reconstruction efficiency}. The efficiency from the signal MC simulation $\varepsilon_{MC}$ is imperfect. The reason is the difficulty of simulating the response of hadronic showers in the detector material due to their complex structure and number of components. As a consequence, the distributions for observables based on the TOF or EMC detector response for MC simulation are not in agreement with the corresponding distributions from the collision data. This leads to an imprecise $\varepsilon_{MC}$ which needs to be corrected. In this analysis we chose to correct the $\varepsilon_{MC}$ with a data-driven method. The determination of the corrected reconstruction efficiency $\varepsilon_{cor}$ is performed individually for each signal event classification category $i =$ A, B, C.\\

To study the efficiency corrections only depending on either the neutron or the anti-neutron selection observables, the two control channels $e^+e^- \to J/\psi \to \bar{p}n\pi^+ $ and $e^+e^- \to J/\psi \to p\bar{n}\pi^- $ are used. These control channels include two charged particles in the final state, which can be used to predict the position of the EMC shower from the neutron and anti-neutron, respectively. This allows us to precisely study the detector impact from a neutron and anti-neutron from data and compare to the corresponding MC simulation. The selection of the control channels follows the discussion in \cite{jpsinnbar}. With the selection of the control channels, a requirement on the recoil momentum $|{\bf p}^{recoil}_{p\pi^-(\bar{p}\pi^+)}| = |{\bf p}_{J/\psi} - {\bf p}_{p(\bar{p})} - {\bf p}_{\pi^-(\pi^+)}|$ is applied to restrict the momentum of the neutron (anti-neutron) from the control channel to the corresponding signal process final state particle momentum $|{\bf p}_{n(\bar{n})}| = \sqrt{(\sqrt{s}/2)^2-m^2_{n(\bar{n})}}$. The category-specific selection criteria for the neutron (anti-neutron) are applied for both, the control sample MC simulation and the data and the corresponding selection efficiencies $\varepsilon^{data}_n,\ \varepsilon^{data}_{\bar{n}},\ \varepsilon^{MC}_{n},\ \varepsilon^{MC}_{\bar{n}}$ are determined. The final efficiency correction $\mathcal{C}_{n\bar{n}}$ is determined as:

\begin{linenomath}
\begin{equation}\label{effnnbar}
\centering
\begin{split}
 \mathcal{C}_{n\bar{n}} =  \sum_{j,k} \mathcal{M}_{j,k} \cdot w_{j,k}, \\
\Delta \mathcal{C}_{n(\bar{n})} = \sqrt{\sum_{j,k} (\Delta\mathcal{M}_{j,k})^2 \cdot w_{j,k,}^2}, \\
\mathcal{M}_{j,k}  = \frac{\epsilon^{data}_{\bar{n}}({\bf p},\cos\theta)\epsilon_{n}^{data}({\bf p},-\cos\theta)}{\epsilon^{MC}_{\bar{n}}({\bf p},\cos\theta)\epsilon_{n}^{MC}({\bf p},-\cos\theta)}
\end{split}
\end{equation}
\end{linenomath}
$w_{j,k}(p,\cos\theta)$ is the normalized distribution in the momentum-position-space from the signal MC simulation after all selection criteria applied. The negative sign of $\cos\theta$ for the neutron efficiencies takes into account the back-to-back behavior of the signal process. The absolute value is determined by using $\Delta\mathcal{M}_{j,k}$ as the individual error in the corresponding bin $j,k$ from the correction matrix, $\mathcal{M}$, and the signal distribution after all selection criteria applied in the corresponding bin, $w_{j,k}$ .\\
The disagreement between the signal MC simulation and data for the selection criterion $E_{extra}$ is studied with the process $e^+e^-\to p \bar{p}$. The process is selected as discussed in \cite{protonscan}. To avoid biases, the selection criterion on $E/{\bf p}$ is replaced by the requirement on the proton EMC shower of $|\cos\theta | < 0.8$. We assume, that the hadronic showers in the EMC from (anti-)neutrons are similar to the ones from (anti-)protons. Using this hypothesis, we study the cut efficiency for $E_{extra}$ from the clean selected sample of $e^+e^-\to p \bar{p}$ events from data and from the signal MC simulation for $e^+e^-\to n \bar{n}$ and determine the efficiency correction $\mathcal{C}_{extra} = \varepsilon^{p\bar{p},data}_{extra}/\varepsilon^{n\bar{n},MC}_{extra}$.\\

{\bf The trigger efficiency correction:} to determine the trigger efficiency for the signal process $e^+e^-\to n \bar{n}$, we study again the process $e^+e^-\to p \bar{p}$ under the discussed hypothesis. In the first step we determine how often the trigger for pure neutral channels is activated in data for the process $e^+e^-\to p \bar{p}$, while pre-selecting the trigger channel for charged tracks (avoiding a correlation between the trigger for the MDC and EMC to prevent a bias). Obtaining the trigger efficiency for the pure neutral final state trigger channel with respect to the deposited energy in the EMC $E^i_{total}$ from event $i$ (quoted from reference \cite{trigger}), we determine the trigger efficiency correction $\mathcal{C}_{trg}$ for our pure neutral final state signal process:\\

\begin{enumerate}[labelsep = .5em, leftmargin = 0pt, itemindent = 3em]
\item the average trigger efficiency $\epsilon_{trg}$ is defined as:
\begin{linenomath}
\centering
\begin{equation}\label{eq:trg}
\begin{split}
\hfill \epsilon_{trg} = {\sum_{bin} \rho(E)_{bin} Trg(E)_{bin}}, \\  Trg(E) = 0.5+0.5 Erf\left(\frac{E- a}{b} \right),\hfill
\end{split}
\end{equation}
\end{linenomath}
where $\rho (E)$ is the normalized, binned, spectrum of the total energy deposition in the EMC from the signal $e^+e^-\to n\bar{n}$.
 $Trg(E)$, a probability that any event will be triggered under the total energy deposition $E$ in the EMC, is obtained with an analysis of $e^+e^-\to p\bar{p}$. To prevent any bias of the EMC, no selection criteria from EMC are used to select $e^-e^+\to p\bar{p}$ and trigger conditions from the MDC and TOF are pre-required to study the EMC response from the control channel. A conditional energy-dependence of the EMC trigger is obtained by comparison between the number of events passing the MDC+TOF+EMC trigger condition with the events which only passing the MDC+TOF trigger conditions (following the approach from Ref.~\cite{trigger}:
\begin{linenomath}
\begin{equation}\label{trg}
\centering
\hfill P_{trg}({\rm EMC}) = \frac{P_{trg}({\rm EMC} + {\rm MDC+TOF})}{P_{trg}({\rm MDC+TOF})},\hfill
\end{equation}
\end{linenomath}
$P_{trigger}({\rm MDC+TOF})$ is close to 1~\cite{trigger}, therefore it is assumed that $P_{trigger}({\rm EMC})$ is a reasonable evaluation of an independent trigger energy dependence. The obtained $P_{trigger}({\rm EMC})$ is fitted with the function $Trg(E)$ (Eqn.~\ref{eq:trg} (right)). The parameters from fit are determined to $a = 0.758\pm0.005$ and $b = 0.334\pm0.009$.
\item $Trg(E)$ is not sensitive to the magnetic field, which is studied by application of different selection conditions. (This conclusion is drawn from the following facts: at BESIII, the magnetic field is 0.9 Tesla and the inner EMC radius is 94 cm (arxiv.org/abs/0911.4960, table 17). As long as a charged proton/antiprotons carry a transverse momentum more than $p_T = 0.25$ GeV/c, it can reach the EMC. Thus choosing $e^+e^- \to p\bar{p}$ events at $\sqrt{s} = 2.125,\ 2.396\ 2.665$ GeV ($\sqrt{s} = 2.125$; $p_T = 0.30$ GeV/c at $\cos\theta = 0.8$) is reasonable to determine the EMC trigger energy dependence $Trg(E)$.)  As described above, the EMC trigger energy dependence $Trg(E)$ is tested with $e^-e^+\to p\bar{p}$ events from different $\sqrt{s}$ samples under different transverse momenta $p_T$ of the proton and it is found that $Trg(E)$ is stable under different $p_T$ and the only difference is the statistical precision. The final $Trg(E)$ is determined at high $p_T$.
\item The correct reconstruction of the total energy deposition in the EMC is crucial for the correct determination of the trigger efficiency. To obtain a reliable total energy deposition distribution $\rho(E)$ from the signal process, the control sample $J/\psi\to p\bar{n}\pi^-$(+c.c. for the similar correction of the neutron impact in the EMC) is used to correct the energy deposition from the (anti-)neutron in the EMC from the signal MC simulation.
\item $Trg(E)$ represents a probability that any event will be triggered under the total energy deposition $E$ in the EMC, independent from particle type of process. Therefore, the obtained EMC trigger energy dependence $Trg(E)$ is re-weighted with the corrected energy spectrum $\rho(E)$ from the signal process $e^-e^+\to n\bar{n}$. This approach is the best available way to reduce potential bias from the deflection of magnetic field and a solution not dependent on the difference between the anti-proton and the anti-neutron annihilation in the detector material.
\end{enumerate}


{\bf Two exclusive corrections in category C:}
\begin{enumerate}[labelsep = .5em, leftmargin = 0pt, itemindent = 3em]
 \item The disagreement between the signal MC simulation and data for the selection criterion $E_{extra}$ is studied with the process $e^+e^-\to p \bar{p}$. The process is selected as discussed in \cite{protonscan}.
To avoid biases, the selection criterion on $E/{\bf p}$ is replaced by the requirement on the proton EMC shower of $|\cos\theta | < 0.8$. The extra energy $E_{extra}$ is defined as the energy deposition in the EMC not coming from $n$ or $\bar{n}$. A cone is constructed around the flight direction of $n(\bar{n})$ with an opening angle of 20(50)$^{\circ}$. The
$n$ and $\bar{n}$ energy deposition in the EMC comes from the hadronic showers of $n$ and the annihilation of $\bar{n}$. Both respective signals are very similar to the hadronic showers of $p$ and the annihilation of $\bar{p}$. The energy deposition due to Bethe-Bloch energy loss can be neglected here. Radiative electromagnetic processes are absent at this $p$/$\bar{p}$ energy. $E_{extra}$ contains all energy deposition in the EMC excluding all energy deposited in the $n(\bar{n})$ cones. Since the control channel $e^{+}e^{-}\to p\bar{p}$ is a similar two particle final state, one can define the cones for $p$ and $\bar{p}$ in the same way as for $e^+e^-\to n\bar{n}$ and the $E_{extra}$ distribution contains the same kind of EMC response (for example from machine background in the EMC, recoiled secondary particles from the $\bar{n}$($\bar{p}$)-annihilation which may be have a large angle with the $\bar{n}$($\bar{p}$) flight direction and are not included in the 50 degree cone, and showers produced by cosmic rays, among others). The reason, why the channel $e^+e^-\to p\bar{p}$ is used for this study is the similarity of the $\bar{n}$ and $\bar{p}$ annihilation. Using this hypothesis, we study the cut efficiency for $E_{extra}$ from the clean selected sample of $e^+e^-\to p \bar{p}$ events from data and from the signal MC simulation for $e^+e^-\to n \bar{n}$ and determine the efficiency correction $\mathcal{C}_{extra} = \varepsilon^{p\bar{p},data}_{extra}/\varepsilon^{n\bar{n},MC}_{extra}$ as illustrated in Fig.~\ref{fig:cee2}.

\begin{figure}[htb]
	\centering
	\includegraphics[width=7.5cm]{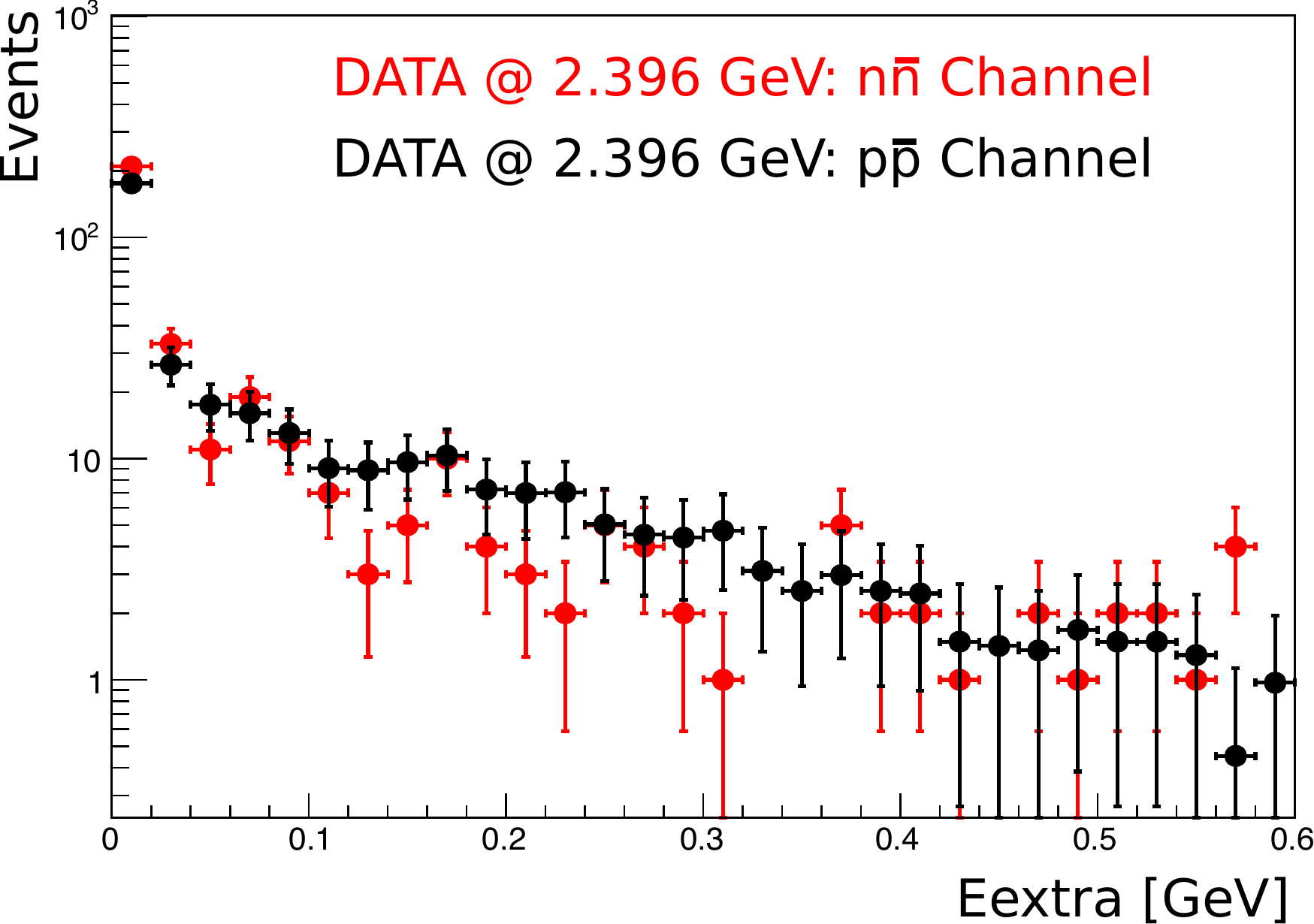}
	\caption{Comparison of E$_{extra}$ distribution from collider data for $e^+e^-\to\bar{p}p$ and $e^+e^-\to\bar{n}n$ selection at $\sqrt{s}=2.396$ GeV for Category C (signal selection as described, additionally a hard cut on the angle between $n$ and $\bar{n}$ is introduced $\sphericalangle^{n}_{\overline{n}} > 175^{\circ}$ to extract the "most signal-like" events). Events in $e^+e^- \to p\bar{p}$ data are scaled to same number of events as in $e^+e^- \to n\bar{n}$ data.}
	\label{fig:cee2}
\end{figure}	

\item
Corrections due to selection criteria on the MUC have been studied in a similar way as for $n$- and $\bar{n}$-based selection criteria. They are found to be negligible and possible residual effects are included in the systematic uncertainty. The corrected reconstruction efficiency $\varepsilon_{cor}$ is the product from the signal MC efficiency $\varepsilon_{MC}$ and the above discussed contributions is:
\begin{linenomath}
\begin{equation}\label{eq:ecor}
\centering
\hfill \varepsilon_{cor} = \varepsilon_{MC}\cdot\prod_i \mathcal{C}_i, (i = {\rm n\bar{n},\ MUC,\ extra,\ trg}) \hfill
\end{equation}
\end{linenomath}
\end{enumerate}

\subsection{  Systematic uncertainties  \label{app:syserror}}
In the first step, the systematic uncertainty on the Born cross section and the effective form factor is determined for each signal classification category. For the final results, the individual systematic uncertainties are combined. The following contributions are studied:
\begin{enumerate}[labelsep = .5em, leftmargin = 0pt, itemindent = 3em]
\item The systematic uncertainty from the luminosity measurement $\delta_L$ is quoted from \cite{luminosity}.

\item The selection criteria for the signal process $e^+e^-\to n\bar{n}$ have been corrected for the difference between the MC simulation and real data using an data-driven approach. We take one standard error of the combined efficiency corrections as the systematic uncertainty due to the signal selection $\delta_{sel}^i$.

\item The uncertainty due to the fit procedure for the extraction of signal event from data combines the contributions from the fitting range, the signal and background shape models. A sum in quadrature of these contributions represents the systematic uncertainty due to the fit $\delta_{fit}^i$.

\item The dependence of the final state particles angular distribution can introduce a systematic effect onto the reconstruction efficiency. To take it into account, we generate two extreme cases for the signal MC simulation samples according to the angular analysis results for $R_{em}^i$, taking into account the corresponding uncertainty. The difference between the signal MC simulation reconstruction efficiency with the nominal signal MC and the two extreme cases is taken as the systematic model uncertainty $\delta_{model}$.

\item The uncertainty from the trigger efficiency $\delta_{trg}^i$ is considered as the difference of the nominal results $\mathcal{C}_{trg}^i$ to results using values from multi-hadronic final states for the parameters $a$ and $b$ in Eq.~(\ref{eq:trg}) instead of the nominal parameters extracted with the $e^+e^-\to p\bar{p}$ process.

\item To estimate the uncertainty from the radiative corrections and the vacuum polarization, we determine the product of the signal MC reconstruction efficiency $\varepsilon_{MC}^i$ and the radiative correction factor $(1+\delta )^i$ for the final and previous form factor parametrization within the signal MC simulation. Additionally, we take into account the parameter uncertainty from the input model for the line-shape via sampling within the uncertainty band. The contributions are taken as the systematic uncertainty $\delta_{ISR}$.

\item Several category-specific systematic uncertainties are considered for non-universal selection criteria. $\delta_{T0}$ and $\delta_{MUC}$ are studied with a data-driven method similar to the efficiency correction study in the previous section, while $\delta_{evt}$ and $\delta_{BDT}$ are studied by variation of the requirements and comparison of outcomes to the nominal results.
\end{enumerate}
For the systematic uncertainty on the Born cross section for one classification category $i = {\rm A,\ B,\ C}$ the contributions are added in quadrature to:

 \begin{equation}\footnotesize
 \label{eqn:sys1}
 \centering
 \begin{split}
\hfill \delta\sigma_{B}^i = \sqrt{\sum_k \delta_k^2},
\end{split}
   \end{equation}
   where {\it k} indicates the labels {\rm $sel$, $fit$, $model$, $trg$, $T_0^{i=A}$, $evt^{i=B}$, $BDT^{i=B}$, $MUC^{i=B,C}$}.
The systematic uncertainty $\delta\sigma_{B}^i$ is propagated as shown in the formula from Eq.~(\ref{formulas}) to determine the corresponding uncertainty on the effective form factor $\delta |G^i|$. The individual systematic uncertainties $\delta\sigma_{B}^i$ are combined to $\delta\sigma_{B}^{comb}$ and $\delta|G^{comb}|$ using the generalized least squares method~\cite{errorweight} from Eq.~(\ref{eqn:weightedXS1}). The uncertainties $\delta_L$ and $\delta_{ISR}$ are considered only once with the expression for the systematic uncertainty $\delta\sigma_{B}$ and $\delta|G|$ on the final results:

 \begin{equation}\label{eqn:sys1a}
 \begin{split}
\hfill \delta\sigma_{B} = \sqrt{(\delta\sigma_{B}^{comb})^2\ + (\delta_L)^2 + (\delta_{ISR})^2}, \\
\delta|G| = \sqrt{(\delta |G^{comb}|)^2 + (\delta_L)^2 + (\delta_{ISR})^2}\hfill
\end{split}
   \end{equation}

\subsection{{\bf Cross-check} \label{app:crosscheck}}

\noindent The signal event classification strategies A, B, and C for the process $e^+e^-\to n\bar{n}$ are tested with the process $e^+e^-\to J/\psi \to n\bar{n}$. We use two data sets which have been collected with the BESIII detector at the c.m.~energy of the $J/\psi$ meson. The total number of the $J/\psi$ events in the data samples is $(223.7\pm 1.4)\times 10^6$ and $(1086.9\pm 6.0)\times 10^6$ \cite{jpsi2012}, respectively.
The number of signal events for $J/\psi\to n\bar{n}$ is determined by fitting $\Delta T_{n}$ or $\sphericalangle^{n}_{\bar{n}}$ as shown in Fig.~\ref{fig:fittingjpsi}.

\begin{figure*}[htb]
  \centering
  \includegraphics[width=5.5cm, height=4.2cm]{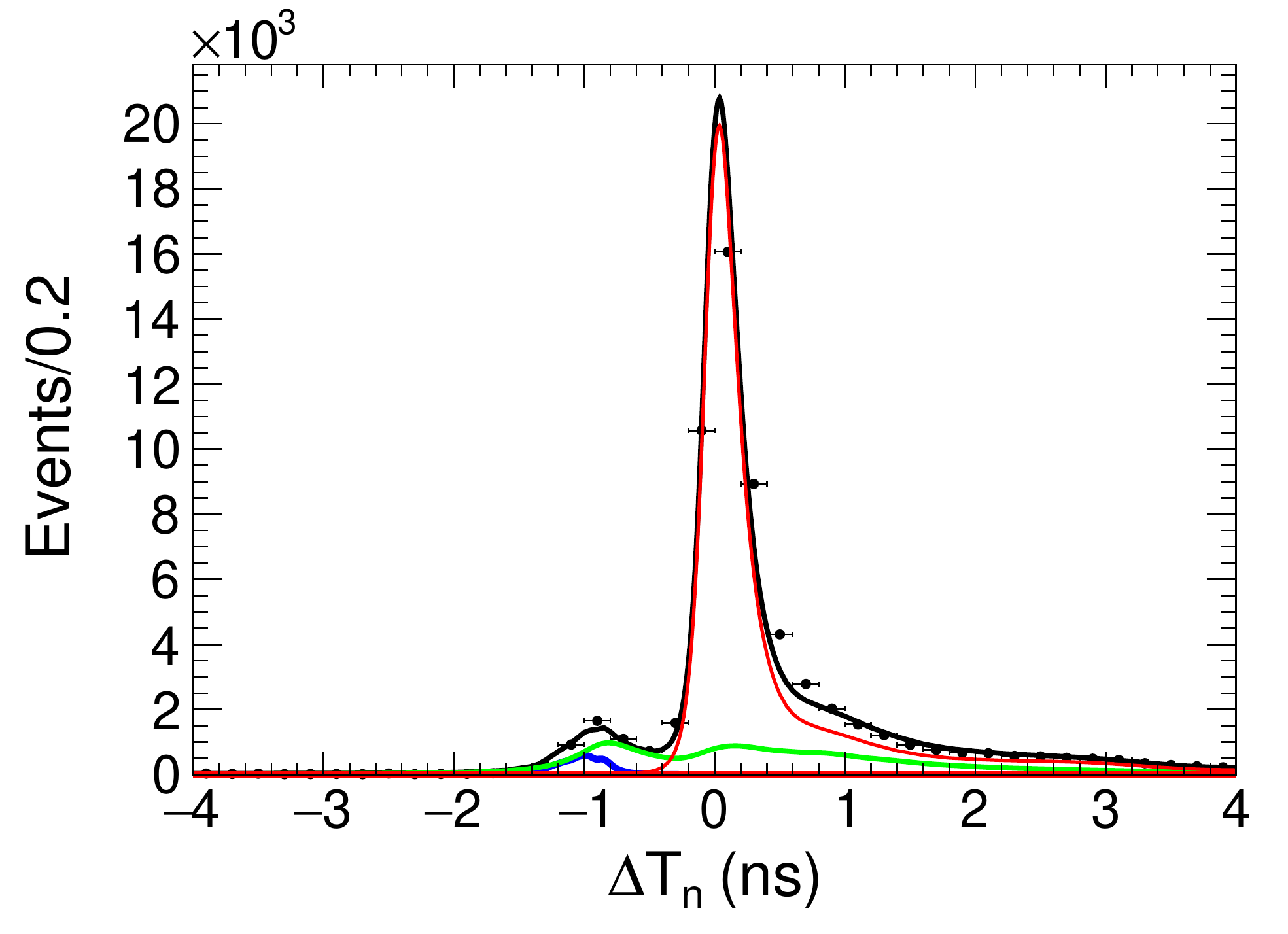}
  \includegraphics[width=5.5cm, height=4.2cm]{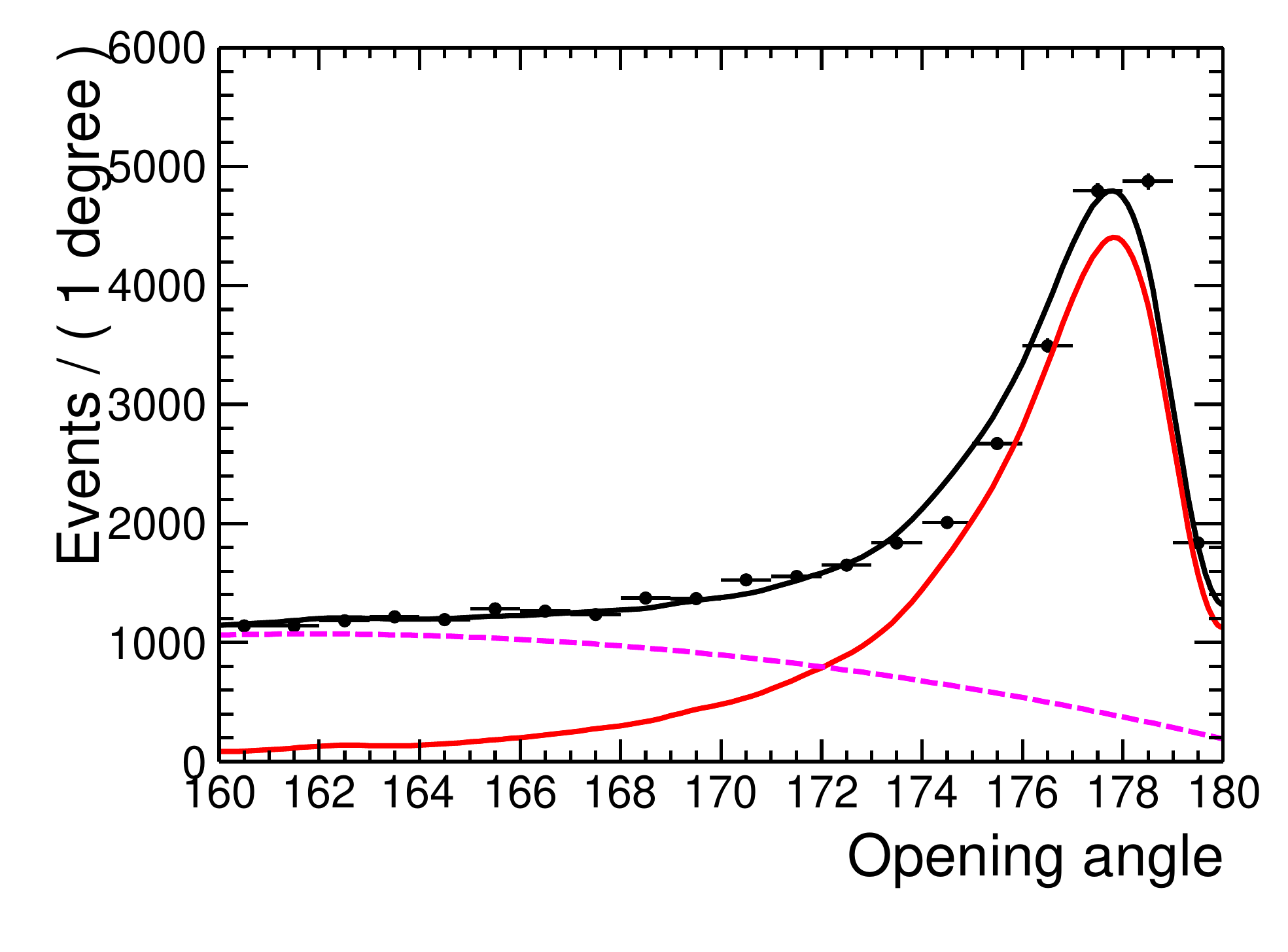}
  \includegraphics[width=5.5cm, height=4.2cm]{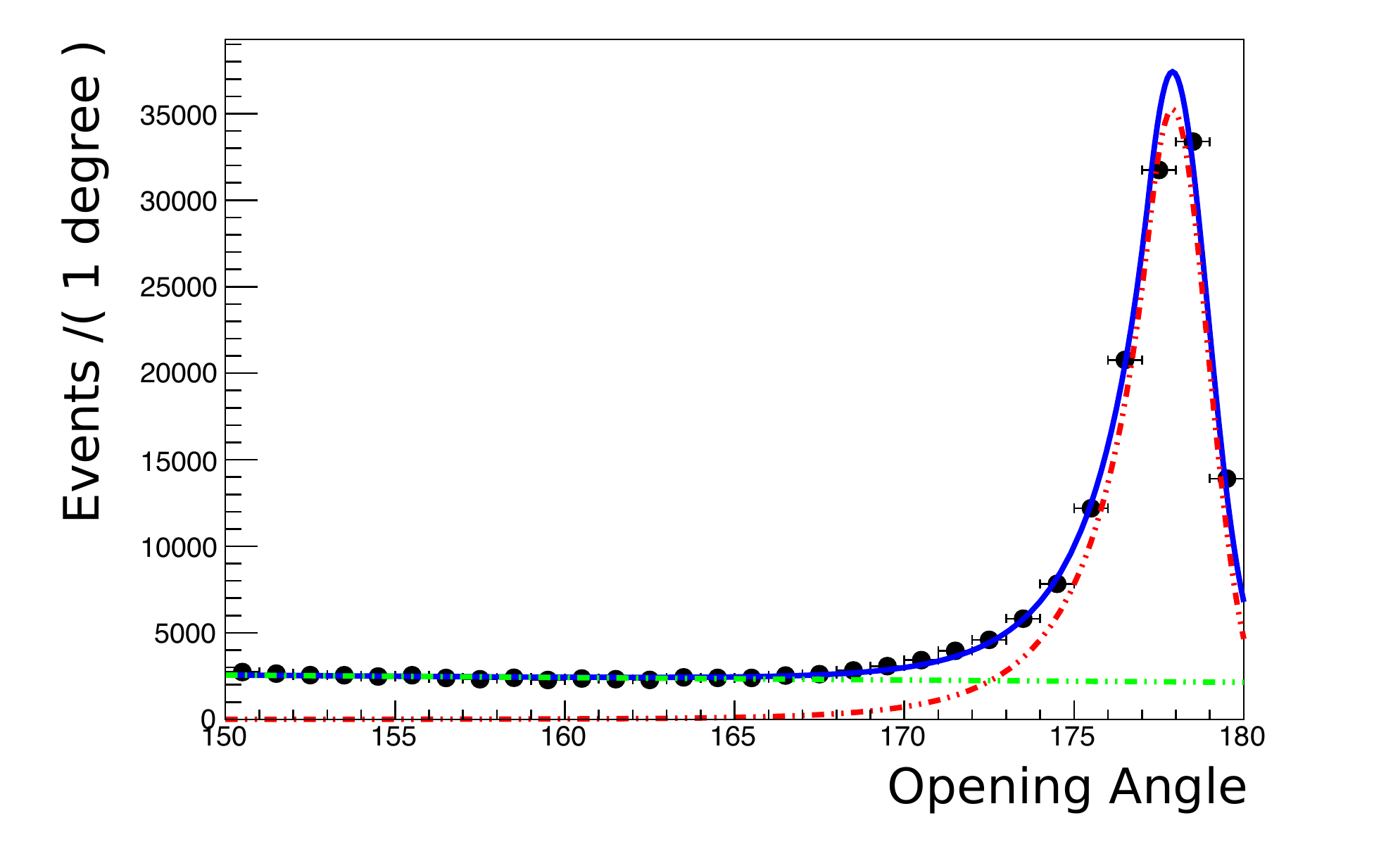}
   \caption{(Left) Category A: determination of the signal events via a fit to the observable $\Delta T_n$ for the selected data at $\sqrt{s} = 3.097$ GeV. Black points represent collider data, the red line represents the signal shape obtained from $J/\psi\to \bar{n}n$ signal MC generated with {\sc conexc}, the green line represents the sideband background, and the blue line represents the $e^+e^-\to\gamma\gamma$ background. The black line represents the combined fit model. (Middle) Category B: determination of the signal events via a fit to the observable $\sphericalangle^{\bar{n}}_n$ for the selected data at $\sqrt{s} = 3.097$ GeV. Black points represent collider data. The red line represents the signal shape obtained from signal MC ($e^+e^-\to J/\psi\to \bar{n}n$), the magenta represents the background shape, the black line represents the combined fit model. (Right) Category C: determination of the signal events via a fit to the observable $\sphericalangle^{\bar{n}}_n$ for the selected data at $\sqrt{s} = 3.097$ GeV. Black points represent collider data, the red line represents the signal shape obtained from signal MC ($e^+e^-\to J/\psi\to \bar{n}n$) generated with {\sc conexc}, the green line represent the sideband background. The blue line represents the combined fit model.}
  \label{fig:fittingjpsi}
\end{figure*}

 The signal MC efficiency is corrected with the control samples $e^+e^-\to J/\psi\to p\bar{n}\pi^-$ and $e^+e^-\to J/\psi\to \bar{p}n\pi^+$, similar to the efficiency correction in the main analysis. We list the branching fractions calculated from data for each category and compare our results to the published reference from~\cite{jpsinnbar}. Our results are consistent with each other and with the world data reference, as shown in Table \ref{tab:jpsi2nnbar}.

\begin{table}[htb]\footnotesize
\begin{center}
\caption{Determination for the branching fraction $B(J/\psi\to n\bar{n})$ [in unit of $\times 10^{-3}$] with the 3 different analysis strategies A, B, and C. The uncertainties are statistical.}
\begin{tabular}{c|c|c|c|c}
 \hline
  \hline
  Category &  A&  B &  C & Ref.~\cite{jpsinnbar}  \\
  \hline
  $B(J/\psi\to n\bar{n})$ & $2.02\pm0.01$ & $2.07\pm0.03$ & $2.00\pm 0.06$  & $2.07\pm0.01$ \\
  \hline
    \hline
\end{tabular}
\label{tab:jpsi2nnbar}
\end{center}
\end{table}


\subsection{{ {\bf The Study of Oscillation in the Effective Form Factor}} \label{app:osc}}

\noindent We study the oscillation of the effective form factor $|G|$ initially for the individual neutron data. Following the approach from~\cite{osz2}, a fit with a modified dipole function $G_{D^{*}} (q^2) \equiv G_{D^*} $ is performed,

\begin{equation}\label{moddip2}
\begin{split}
\hfill G_{osc}(q^2) = |G_{n}| - G_{D^*}, \\
 G_{D^*} = G_{D}\cdot  \frac{1}{1+\frac{q^2}{m_a^2}},  \\
 G_D = \frac{\mathcal{A}_{n}}{ \left( 1-\frac{q^2}{0.71 ( \textrm{GeV}^2)}  \right)^2 },
\end{split}
\end{equation}
and the parameters determined to $\mathcal{A}_n = 3.5\pm0.1$, $m^2_a\sim50\times 10^{6}$. The very large parameter for $m^2_a$ indicates that our data can be described by the common dipole formula $G_{D}$ with the normalization $\mathcal{A}_{n} = 3.5\pm0.1$. Figure \ref{fig:indneutronpara}(a) shows the two different parameterizations. The two different parameterizations are shown as the blue and red dotted line overlaying each other.

To describe the oscillation of the reduced form factor $G_{osc}$, a fit with Eq.~(\ref{funcosz}) (+ an additional Pol0) is performed. The parameterization can describe our data well with $\chi^2/ndf = 11.8/13$, as shown in Fig.~\ref{fig:indneutronpara}(b). We obtain the parameters for the momentum frequency $C^n = 5.28\pm0.36$ GeV$^{-1}$ in comparison to the proton frequency of $C^p = 5.5\pm0.2$ GeV$^{-1}$ from \cite{osz2} and to the shared frequency from the simultaneous fit to the nucleon data with $C = 5.55\pm0.28$ GeV$^{-1}$.

To investigate the significance of the oscillation, we compare the fit with the Eq.~(\ref{funcosz}) for the description of oscillation $F_{osc}$ with an additional possible shift ($F_{osc}$ + Pol0) to a fit with a polynomial of zeroth degree, as shown in Fig.~\ref{fig:indneutronpara}(b). The statistical significance of the periodic structure is 6.3$\sigma$.

\begin{figure*}[htb]
	\centering
     \begin{overpic}[width=0.49\textwidth]{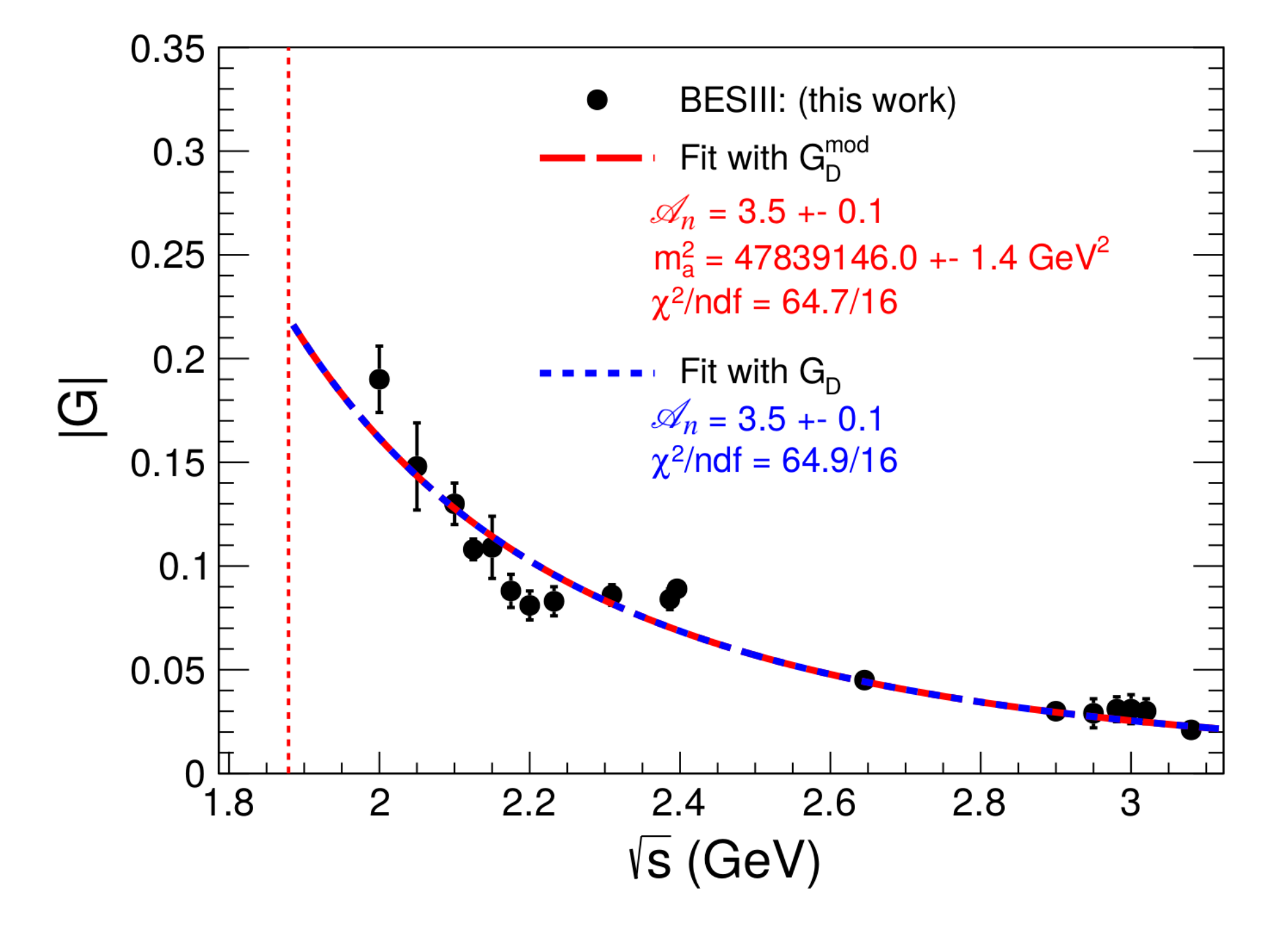}
     \put(-2, 66) {{\bf \Large{(a)}}\color{black}}
     \end{overpic}
     \begin{overpic}[width=0.49\textwidth]{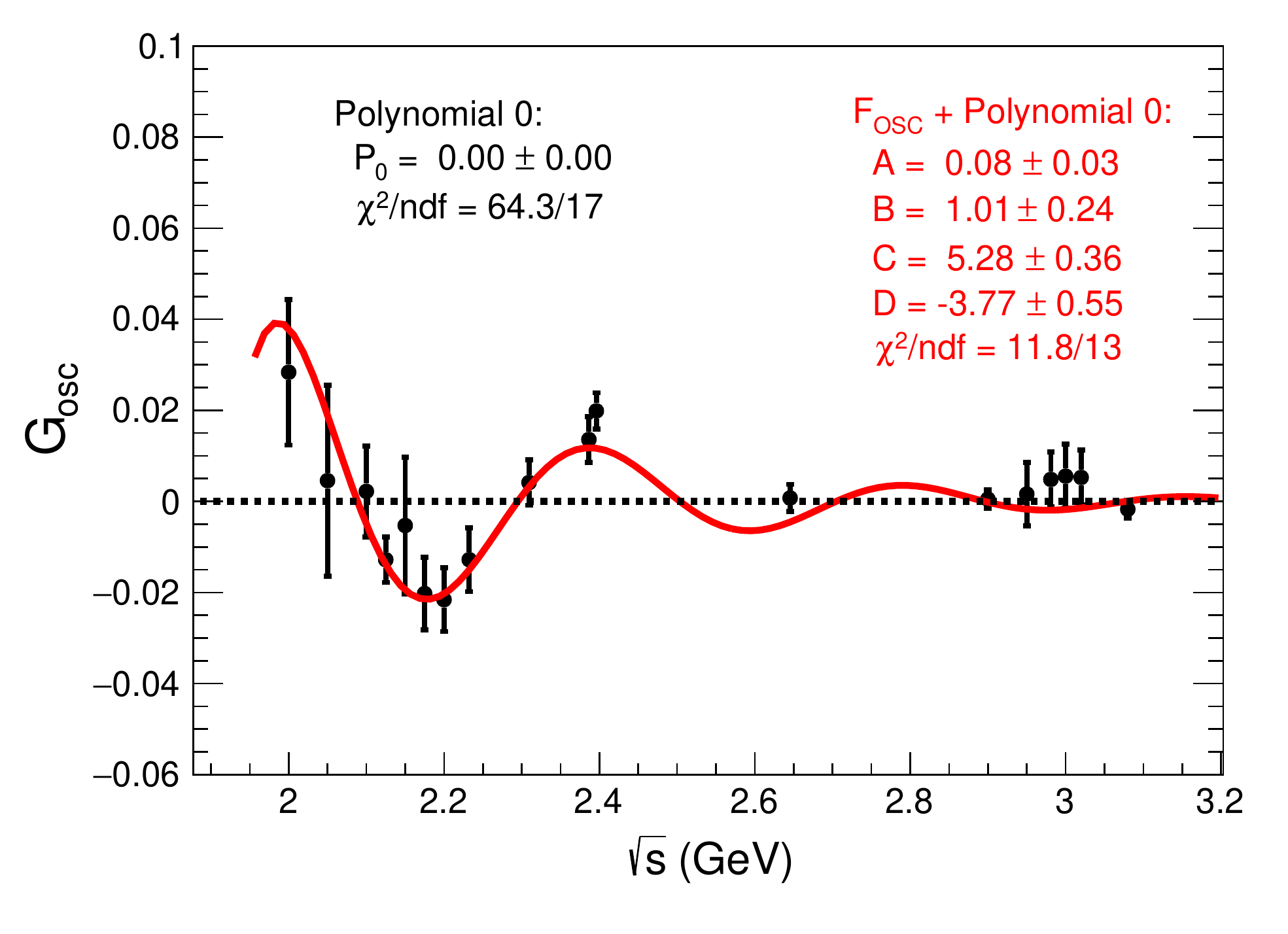}
      \put(-2, 66) {{\bf \Large{(b)}}\color{black}}
     \end{overpic}
	\caption{(a) Fit to the effective form factor of the neutron $|G|$. The blue and red dotted lines are fits with the functions \ref{moddip} and \ref{moddip2}. (b) Fit with Eq.~(\ref{funcosz}) to $G_{osc}$ using the parametrization \ref{moddip} shown as the red solid line. The statistical significance for the oscillation is carried out by comparison to a fit with a polynomial of zeroth degree (black dotted line), and is 6.3$\sigma$. The obtained parameters are shown in the plots.}
	\label{fig:indneutronpara}
\end{figure*}

\end{document}